\documentclass[11pt]{article}
\usepackage{amsmath,amsfonts,amsthm,amssymb}

\newcommand{\version}{November 22, 2005}

\font\notefont=cmsl8  
\newcommand\x{x}
\newcommand\bfx{{\bf x}}
\newcommand\y{y}
\newcommand\z{z}
\newcommand\p{p}
\newcommand\vecA{A}

\newcommand\ham{{\mathcal K}}
\def\an#1{a_{#1}^{\phantom{\dagger}}}
\def\ad#1{a_{#1}^\dagger}
\newcommand\vecz{{\bf z}}
\newcommand\pbos{P_{\rm b}}
\newcommand\infspec{{\rm inf\, spec\, }}
\newcommand\E{{\mathcal E}}

\newcommand\Egp{{\mathcal E}^{\rm GP}}
\newcommand\Engp{E^{\rm GP}}

\newcommand\C{{\mathbb C}}

\newcommand\R{{\mathbb R}}
\newcommand\eps{\varepsilon}
\newcommand\half{\mbox{$\frac 12$}}
\renewcommand\rho\varrho

\newcommand\const{{\rm const.\, }}
\newcommand\Tr{{\rm Tr}}

\newcommand{\F}{{\cal F}}

\newcommand\Bb{{\cal B}}

\newcommand\id{{\mathbb I}}

\newcommand{\K}{K}
\newcommand{\M}{\widetilde{M}}

\newcommand{\s}{S}
\def\limi{\mathop{\rm \underline {lim}}\limits}

\newtheorem{thm}{Theorem}
\newtheorem{lem}{Lemma}
\newtheorem{prop}{Proposition}

\theoremstyle{definition}

\begin{document}

\title{Derivation of the Gross-Pitaevskii Equation\\ for Rotating Bose Gases}
\author{\vspace{5pt} Elliott H.~Lieb$^{*}$ \\
\vspace{-4pt}\small{ Departments of Mathematics and Physics,
Jadwin Hall,} \\
\small{Princeton University, P.~O.~Box 708, Princeton, New Jersey
  08544}
\and Robert Seiringer$^{\dagger}$\\
\vspace{-4pt}\small{ Department of Physics,
Jadwin Hall,} \\
\small{Princeton University, P.~O.~Box 708, Princeton, New Jersey
  08544}
}
\date{\small \version}
\maketitle

\renewcommand{\thefootnote}{$*$}
\footnotetext{Work partially
supported by U.S. National Science Foundation
grant PHY 01 39984.}
\renewcommand{\thefootnote}{$\dagger$}
\footnotetext{Work partially
supported by U.S. National Science Foundation
grant PHY 03 53181, and by an A.P. Sloan Fellowship\\
\copyright\, 2005 by the authors. This paper may be reproduced, in its
entirety, for non-commercial purposes.}

\begin{abstract}
  We prove that the Gross-Pitaevskii equation correctly describes the
  ground state energy and corresponding one-particle density matrix of
  rotating, dilute, trapped Bose gases with repulsive two-body
  interactions.  We also show that there is 100\% Bose-Einstein
  condensation. While a proof that the GP equation correctly describes
  non-rotating or slowly rotating gases 
  was known for some time, the rapidly rotating case was unclear
  because the Bose (i.e., symmetric) ground state is {\it not\/} the
  lowest eigenstate of the Hamiltonian in this case. We have been able
  to overcome this difficulty with the aid of coherent states. Our
  proof also conceptually simplifies the previous proof for the slowly
  rotating case. In the case of axially symmetric traps, our results
  show that the appearance of quantized vortices causes
  spontaneous symmetry breaking in the ground state.
\end{abstract}

\bigskip
\centerline{
{\it \large Dedicated to Jakob Yngvason on the occasion of his 60$^{\it th}$ birthday}}

\section{Introduction}

In this paper we show that a dilute, rotating Bose gas is correctly
described by the Gross-Pitaevskii (GP) equation. We also show that
there is 100\% Bose-Einstein condensation (BEC) into a solution of
the GP equation in a suitable limit. These conclusions were heretofore
unproved, and it might not be an exaggeration to say that they were
even conjectural, primarily because of the unusual situation (proved
in \cite{S03}) that the absolute ground state of the
Schr\"odinger Hamiltonian is {\it not the bosonic ground state\/} in the
rapidly rotating case, as it is in the case when there is little or no
rotation. In other words, the vortices seen in rotating gases are not
properties of the absolute ground state but are, instead, true
manifestations of the bosonic symmetry requirement. 

If the GP equation correctly describes the physics of a rotating gas
(as we show here), then it also shows the superfluidity of such a gas,
as will be discussed below. In the case of a cylindrically symmetric
trap potential, the rotational symmetry is broken when more than one
vortex is present; the GP equation must describe this broken
rotational symmetry and, therefore, it must have multiple minimum
energy solutions in this case.

The key mathematical tool employed here is coherent states. Our work
is based on the results of \cite{LSYjust} and the observation there
that one can make a $c$-number substitution for many boson modes (not
just one, as in Bogoliubov's method) without significant error
provided the number of such modes is of lower order than $N$, the
number of particles.

As in our previous work \cite{LSY00,LS,LSYsuperflu,S03,book} on dilute, 
trapped Bose gases we start with the 
Hamiltonian for $N$ bosons
\begin{equation} \label{ham}
H_{\!_N} = \sum_{i=1}^N H_0^{(i)} + \sum_{1\leq i<j\leq N} v_{\!_N}(\x_i-\x_j)\,,
\end{equation}
where $H_0$ is the one-body part of the Hamiltonian and $v_{\!_N}$ is the
two-body repulsive interaction. These terms and the GP limit are
described as follows.

{\bf 1. The GP limit:} We want to fix the external trapping potential
but let $N$ tend to infinity. To retain the notion of a dilute gas in
this situation we let the interparticle potential depend on $N$ in
such a way that $a_{\!_N}$, the two-body scattering length of
$v_{\!_N}$, is related to $N$ by the condition that
\begin{equation}
Na_{\!_N} = a \quad\quad\quad {\mathrm{ is \ fixed.}}
\end{equation} 
In this limit the three components of the energy (kinetic, trapping
potential and interaction potential) scale in the same way and are all
of the same order of magnitude. We call this the {\it GP limit}.
It is this limit that will lead to the GP equation (\ref{gpequation}).

{\bf 2. The two-body potential:} We choose a radial two-body potential
$w(x)$ such that $w(x) \geq 0$ (this is an important restriction for
our methods) and such that $w(x) =0 $ for $|x|> R_0$ (this finite
range condition is a technical restriction for simplicity and can be
relaxed if need be). We note that integrability of $w(x)$ is not
assumed here, $w(\x)$ is even allowed to have a hard core.  The
scattering length of $w$ is $a$ (i.e., the solution to
$[-\frac{1}{2}\Delta +w(x)]f(x)=0 $ with $f(\infty)=1$ satisfies
$f=1-a/|x|$ for $|x|>R_0$). The actual two-body potential in
(\ref{ham}), given by 
\begin{equation}
v_{\!_N}(\x) = N^2 w(N \x) \, ,
\end{equation}
has scattering length $a_{\!_N}= a/N$.

{\bf 3. The one-body Hamiltonian:} We work, as usual, in the rotating
coordinate system, in which case the kinetic energy has to be
supplemented by a term $-\Omega\cdot(p\wedge \x) = p\cdot (\Omega
\wedge x)$, where $\Omega $ is the angular velocity vector, and
$p=-i\hbar\nabla$. It is convenient to add and subtract a
term $\frac{m}{2} (\Omega \wedge \x)^2$ and thereby write
\begin{equation}\label{obham}
H_0 = \frac{1}{2m} (p + \vecA(\x) )^2 + V(x)
\end{equation}
with $\vecA(\x) = m \Omega\wedge \x$. Then $V$ is the trapping
potential (which might or might not have some geometric symmetry) {\it
  minus\/} $\frac{m}{2} (\Omega\wedge \x)^2$. It is well known that we
must have $V(x)\to \infty$ as $|x|\to \infty$, for otherwise the
system will fly apart. We can also assume that $V\geq 0$ without loss
of generality. Actually, for technical reasons we require just a
little more, namely $V(\x) \geq C_1 \ln(|x|) - C_2$ for some positive
constants $C_1$ and $C_2$. (This condition can probably be relaxed a
bit. What we actually need is that $\Tr\, e^{\alpha(\Delta-V(x))}$ and
$\Tr\, |\vecA(\x)|^s e^{\alpha(\Delta-V(x))}$ are finite for $\alpha$
large enough, for some $s>2$. We will show in the appendix that this
is fulfilled under the stated assumption on $V$.)
\medskip

We note that in the rotating coordinate system, the velocity at $x$ is
not $p/m$ but rather $v=i\hbar^{-1} [H_0,x]=p/m + \Omega \wedge x$.
The angular velocity around the $\Omega$ axis is $\Omega \cdot
(v\wedge x) |x^\perp|^{-2} |\Omega|^{-1}$, where $|x^\perp|$ is the
distance to the $\Omega$ axis. In a cylindrically symmetric state
$\psi$ we have $\Omega\cdot(p\wedge x)\, \psi=0$ and, therefore, the
angular velocity is $\Omega$, not zero. In the {\it fixed} frame the
angular velocity is $\Omega-\Omega=0$. In other words, the system in
such a state is not rotating. As long as $\Omega$ is small enough, the
GP ground state is cylindrically symmetric and hence there is no
rotation; this is a manifestation of superfluidity. In order to have
rotation at least one vortex must form. This is a typical property of
superfluids.

\medskip
Henceforth, we use units in which $\hbar = 2m =1$. We also note that
the modification of the kinetic energy in (\ref{obham}) is
mathematically just like that caused by a uniform magnetic field with
vector potential $A$ (and $e/c =1$). There is nothing special about
$\vecA(\x) = m \Omega\wedge
\x$ as far as the mathematics is concerned, so one could have an
arbitrary $\vecA$ without disturbing our analysis, provided it did not
grow too fast at infinity.  One could think, for example, of applying
a magnetic field to the system, but then our particles would have to
be charged and the attendant Coulomb interaction would nullify the
treatment of the system as a dilute gas with short range interaction.
On the other hand, we could allow our particles to have a magnetic
moment (``bosons with spin'') and our analysis would easily extend to
this case. The ground state energy depends in a non-trivial way on the
total spin when there is rotation \cite{S02}, even in the absence of a
magnetic field. This is due to the symmetry requirement of the wave
function, whereby the symmetry of the spin part determines the spatial
symmetry (see, e.g., \cite{eisenberg}).  We will not pursue this topic
further in this paper.

Our analysis is carried out here for three-dimensional gas particles,
but the same ideas apply to a two-dimensional gas. There will be
changes, of course, because the notion of scattering length is
different in 2D and because the energy per particle of a homogeneous
gas of low density $\rho$ is not $4\pi \rho a_s$ as in 3D but rather
$4\pi \rho / |\log \rho a_s^2|$. (Here, $a_s$ is the unscaled
scattering length of the interaction potential, which is held fixed in
the thermodynamic limit for the homogeneous gas.) Thus, the GP
equation will be a little different, but the conclusion will be the
same: The only effect of rotation is to replace $p^2$ by $|p+A|^2$ in
the GP equation derived in \cite{2dgp}.  In order to keep this paper
manageable we do not discuss the 2D case, but the interested reader
can easily combine the results in \cite{2dgp}, \cite{S03} and the
present paper.

The Hamiltonian $H_{\!_N}$ acts on $L^2(\R^{3N})$ but we are
interested in its restriction to the {\it bosonic\/} subspace of
$L^2(\R^{3N})$, namely to permutation symmetric functions. We denote
the ground state energy of $H_{\!_N}$ in the bosonic sector by
$E_0(N)$, and we keep in mind that this might be larger than the
absolute ground state energy of $H_{\!_N}$ when no permutation
symmetry is imposed.

\medskip
We turn now to the GP equation, which originates from the GP energy
functional for a complex-valued function $\phi$ of one variable $x\in \R^3$. 
For $a\geq 0$, the GP energy functional is given by
\begin{equation}\label{gpenergy}
\Egp[\phi] = \langle \phi| H_0 |\phi\rangle + 4\pi a \int_{\R^3} |\phi(\x)|^4 d\x \ .
\end{equation}
It can easily be shown \cite{LSY00} that $ \Egp[\phi]$ has a minimum
over all $\phi$ with $\|\phi\|_2=1$ and this minimum energy is denoted
by $\Engp(a)$.  (We use the standard notation $\|\phi\|_p = \left[\int
  |\phi(x)|^p\, dx \right]^{1/p}$.)  There might be several minimizers (and
there surely will be when the trap has axial symmetry and $a$ is
large \cite{S02,S03}) but each minimizing $\phi$ will satisfy the GP equation
\begin{equation}\label{gpequation}
(-i\nabla +\vecA(x))^2 \phi(x) +V(x) \phi(x) +8\pi a |\phi(x)|^2 \phi(x) = 
\mu \phi(x) \, ,
\end{equation}
where $\mu$ is the chemical potential (i.e., the energy per particle
to add a small number of particles).  Note that $\mu = \Engp(a) + 4\pi
a \int |\phi(\x)|^4 d\x > \Engp(a)$ because of the quartic
nonlinearity.

Our main theorem concerning the bosonic ground state energy of
(\ref{ham}) is the following.

\begin{thm}\label{energy}
  With $a$ denoting the scattering length of $w$, we have
\begin{equation}\label{thmeq}
\lim_{N\to\infty} \frac {E_0(N)}{N} = \Engp(a) \,.
\end{equation}
\end{thm}

In \cite{S03} it was shown that
\begin{equation}\label{upperbound}
\limsup_{N\to\infty} \frac {E_0(N)}{N} \leq \Engp(a) \, ,
\end{equation}
and, therefore, it remains only to prove a lower bound to $\liminf_{N\to\infty} 
{E_0(N)/N}$ of the right form, which we do here.

The GP energy minimizer(s) $\phi$ also tells us something about the
density (diagonal and off-diagonal) and about Bose-Einstein
condensation in the ground state of $H_{\!_N}$, or any \underline{\it approximate
  ground state}. We call a {\it sequence\/} of bosonic $N$-particle density
matrices $\gamma_{\!_N}$ an approximate ground state if
$\lim_{N\to\infty} N^{-1} \Tr\, H_{\!_N}\gamma_{\!_N} = E^{\rm GP}(a)$.
The reduced one-particle density matrix of $\gamma_{\!_N}$ will be
denoted by $\gamma_{\!_N}^{(1)}$.

We would like to suppose that, as $N\to\infty$, $\gamma_{\!_N}^{(1)}$
converges to some $\gamma$ and that $\gamma = |\phi\rangle\langle
\phi|$, where $\phi$ is a solution to the GP equation. This
would be 100\% Bose-Einstein condensation into the GP state and was
proved to occur in the non-rotating case \cite{LS}. The difficulty in
the rotating case is that the solution to the GP equation might not be
unique (as it is in the non-rotating case), in which case the limit
$\gamma$ need not be a pure state. We would expect, however, that
$\gamma $ is always a convex combination of pure GP states, i.e.,
$\gamma = \sum_i \lambda_i |\phi_i\rangle\langle \phi_i|$, where
$\phi_i$ is a solution to the GP equation and $ \sum_i \lambda_i =1$.
(This, of course, is not the same as the much weaker and less
interesting statement that $\gamma$ is a convex combination of terms
of the form $|\psi \rangle \langle \psi |$, in which $ \psi $ is a
linear combination of GP solutions instead of being equal to just one
GP solution.)  Unfortunately, as in the case of a cylindrically
symmetric trap, the set of GP states might not be countable, and so
the summation $\sum_i$ must be replaced by some kind of integral. This
accounts for the rather abstract Theorem \ref{condensation} below. In
any event, this theorem tells us that there is always 100\%
condensation, even if the system has a wide choice of states into
which to condense.

Note that $ \gamma_{\!_N}^{(1)}$ is a positive trace class operator on
the one-particle space $L^2(\R^3)$, and we choose the normalization
$\Tr\, \gamma_{\!_N}^{(1)}=1$ for convenience. (The conventional
normalization is $\Tr\,\gamma_{\!_N}^{(1)}=N$.) By the Banach-Alaoglu
Theorem, any sequence $\gamma_{\!_N}^{(1)}$ will have a subsequence
that converges to some $\gamma$ in the weak-* topology, i.e.,
$\lim_{N\to\infty} \Tr\, A\gamma_{\!_N}^{(1)}= \Tr\, A\gamma$ for all
compact operators $A$. This convergence will even hold in the norm
topology, i.e., $\lim_{N\to\infty} \Tr\,
|\gamma_{\!_N}^{(1)}-\gamma|=0$ by compactness.  More precisely, since
the $\gamma_{\!_N}^{(1)}$ are the one-particle density matrices of
approximate ground states, we have (using the positivity of the
interaction potential in $H_{\!_N}$) $\Tr\, H_0
\gamma_{\!_N}^{(1)}\leq \const$ independently of $N$. Hence also
$\sqrt{H_0}\gamma_{\!_N}^{(1)}\sqrt{H_0} \rightharpoonup \sqrt{H_0}
\gamma \sqrt{H_0}$ in weak-* sense, i.e., $\Tr\, A \sqrt{H_0}
\gamma_{\!_N}^{(1)}\sqrt{H_0} \to \Tr\, A \sqrt{H_0} \gamma
\sqrt{H_0}$ for all compact $A$. Since $H_0^{-1}$ is a compact
operator, this implies that $\Tr\, \gamma_{\!_N}^{(1)} \to \Tr\,\gamma$
as $N\to \infty$ (simply use $A=H_0^{-1}$ above). For positive
operators, weak-* convergence plus convergence of the trace implies
norm-convergence~\cite{Wehrl,Simon}.

We denote by $\Gamma$ the set of all $\gamma's$ that are limit points
of one-particle density matrices of approximate minimizers. That is,
\begin{equation}\label{defgamma}
\Gamma= \left\{ \gamma\, : \, \mathrm{there\ is\ a} {\rm \ sequence\ } \gamma_{\!_N}, 
\lim_{N\to\infty} \frac 1N \Tr\, H_{\!_N} \gamma_{\!_N}= E^{\rm GP}(a),\ 
\lim_{N\to\infty} \gamma_{\!_N}^{(1)}=\gamma\right\}\,.
\end{equation}
As remarked above, the convergence $\gamma_{\!_N}^{(1)}\to \gamma$ can
either mean weak-* convergence or norm convergence. Note that, in
particular, norm convergence implies that $\Tr\,\gamma=1$ for all
$\gamma\in \Gamma$.

\begin{thm}  \label{condensation}
  The set $\Gamma$ of one-particle density matrices of approximate
  ground states, as defined in (\ref{defgamma}), has the following
  properties.
\begin{itemize}
\item[(i)] $\Gamma$ is a compact and convex subset of the set of all
  trace class operators.
\item[(ii)] Let $\Gamma_{\rm ext}\subset \Gamma$ denote the set of
  extreme points in $\Gamma$. (An element $\gamma\in \Gamma$ is
  \underline{extreme} if $\gamma$ cannot be written as $\gamma =
  a\gamma_{_1}+ (1-a)\gamma_{_2}$ with $\gamma_{_{1,2}}\in \Gamma$,
  $\gamma_{_1}\neq \gamma_{_2}$, and $0<a<1$.)  We have $\Gamma_{\rm
    ext} = \{ |\phi\rangle\langle\phi|\, : \, \Egp[\phi]=E^{\rm
    GP}(a)\}$, i.e., the extreme points in $\Gamma$ are given by the
  rank-one projections onto GP minimizers.
\item[(iii)] For each $\gamma\in \Gamma$, there is a positive (regular Borel) measure
  $d\mu_\gamma$, supported in $\Gamma_{\rm ext}$, with
  $\int_{\Gamma_{\rm ext}} d\mu_\gamma(\phi) =1$, such that
\begin{equation}\label{3eq}
\gamma = \int_{\Gamma_{\rm ext}} d\mu_\gamma(\phi)\, |\phi\rangle\langle\phi| \,,
\end{equation}
where the integral is understood in the weak sense. That is, every
$\gamma\in\Gamma$ is a convex combination of rank-one projections onto
GP minimizers.
\end{itemize}
\end{thm}

A consequence of the Krein--Milman Theorem \cite[vol. 2, Thm. 25.12]{choq} is
that given any $\gamma \in \Gamma$ and given any $\varepsilon>0$ there
are  finitely many GP minimizers $\phi_i$ and positive coefficients
$\lambda_i$ (with $\sum_i \lambda_i=1$) such that
\begin{equation}\label{finite}
\gamma = \sum_i\, \lambda_i\, |\phi_i\rangle \langle \phi_i| + \Delta_\varepsilon
\end{equation}
with $\Tr \, |\Delta_\varepsilon | <\varepsilon$.  That is, every
element of $\Gamma$ can be approximated by a {\it finite} convex
combination of GP minimizers.  We also note that part (iii) of
Theorem~\ref{condensation} follows from part (ii) using Choquet's
Theorem \cite[vol. 2, Thm. 27.6]{choq}. We shall, however, prove part
(iii) (and Eq. (\ref{finite})) directly in Section 3 (see Step~4).

\medskip Eq. (\ref{3eq}) reflects the {\it spontaneous symmetry
  breaking\/} that occurs in the system under consideration. Consider
the case of an external potential $V(x)$ which is axially symmetric,
with symmetry axis given by the angular velocity vector $\Omega$. In
general, the non-uniqueness of the GP minimizer stems from the
appearance of {\it quantized vortices}, which break the axial
symmetry, and hence lead to a whole continuum of GP minimizers
\cite{S02,S03,BR99,CD99,GP99,FS01,AD01}.  Uniqueness of the GP
minimizer can be restored by perturbing the one-particle Hamiltonian
$H_0$ in such a way as to break the symmetry and to favor one of the
minimizers, e.g., by introducing a slightly asymmetric trap potential
$V(x)$.  This then leads to {\it complete BEC}, as can be seen from
our Theorem~\ref{condensation}, which does not assume any particular
symmetry of $V(x)$. Note that in the case of a unique GP minimizer,
Theorem~\ref{condensation} implies that the reduced one-particle
density matrix of {\it any\/} approximate ground state converges to
the projection onto this unique GP minimizer, since $\Gamma_{\rm ext}$
(and hence $\Gamma$) consists of only one element in this case.

\medskip
The situation of a dilute rotating Bose gas described in this section
contrasts with the situation of the {\it absolute\/} ground state of
$H_{\!_N}$, i.e., the lowest eigenvalue and corresponding state without
imposing symmetry restrictions on the wavefunctions. In \cite{S03} it
was shown that Eq. (\ref{thmeq}) does  {\it not\/} hold, in general, for
the absolute ground state energy. The energy per particle in this case
is given by minimizing a functional similar to (\ref{gpenergy}), but
which now depends on one-particle density matrices rather than on wave
functions $\phi(x)$. In \cite{S02,S03} it was shown that the
corresponding energy is {\it strictly lower\/} than $\Engp(a)$ for $a$
large enough (and $\Omega\neq 0$). The density matrix functional has a
{\it unique\/} minimizer for any value of $\Omega$ and $a$, and in
general this minimizer will not be rank one. An analogue of
Theorem~\ref{condensation} also holds for the absolute ground state.
As shown in \cite{S03}, $\Gamma$ consists of only one element in this
case, namely the unique minimizer of the density matrix functional
just mentioned. This implies, in particular, that there is no spontaneous
symmetry breaking in the absolute ground state. We refer the reader to
\cite{S03} for more details.

In the remainder of this paper, we present the proof of
Theorems~\ref{energy} and~\ref{condensation}.

\bigskip {\it Acknowledgment.} We are grateful to Lev Pitaevskii for
drawing our attention to the problem of the correctness of the GP
equation for a rapidly rotating Bose gas in an email correspondence in
1999.

\section{Proof of Theorem \ref{energy}} \label{sec2}


\subsection*{STEP 1. Reduction of the Number of Particles to Ensure a Bounded Energy per Particle.}  

One of the problems we shall face in our analysis is to control
three-body collisions, i.e., to show that the ground state wave
function is suitably small when three particles are close together. We
have found a way to do this (see Step 4) with the help of a bound on
the change in energy when three particles are added to the system. It
is not evident that this bound is always satisfied (although it must
be satisfied on average since the total energy is bounded by $N$) and
the discussion in this subsection shows how to circumvent this
annoyance. If another way could be found to control the three-body
amplitude or to control the incremental energy then the analysis in
this section would not be needed.

Let us consider the Hamiltonian (\ref{ham}) for $M\leq N$ particles
(but still with interaction potential $v_{\!_N}$ depending on $N$):
\begin{equation}\label{ham1}
H_{\!_{M,N}} = \sum_{i=1}^M H_0^{(i)} + \sum_{1\leq i<j\leq M} v_{\!_N}(\x_i-\x_j)\,.
\end{equation}
This operator acts naturally on all of $L^2(\R^{3M})$. We denote the
ground state energy in the {\it bosonic\/} sector by $E_0(M,N)$. Our goal is 
a good lower bound on $E_0(N,N)$. 

Let $\M=\M(N)$ be the largest integer $\leq N$ satisfying two
conditions: a.) $N-\M$ is divisible by 3 and b.)
$E_0(\M,N)-E_0(\M-3,N)\leq 6\Engp(a)$. Then $E(\M+3,N) -E(\M,N)>
6\Engp(a),\ E(\M+6,N) -E(\M+3,N)> 6\Engp(a),$ etc., whence
\begin{equation}\label{MN}
E_0(N,N) \geq E_0(\M,N) + 2(N-\M) \Engp(a) \ .
\end{equation}
We will prove the following in the remainder of this section.

\begin{prop} \label{mainlemma}
  Fix $Z>0$, and let $M_j$ and $N_j$ be two sequences of integers, with $M_j\leq
  N_j$, $\lim_{j\to\infty} M_j=\infty$ and $\lim_{j\to\infty}
  N_j=\infty$, such that $E_0(M_j,N_j)-E_0(M_j-3,N_j)\leq 3Z$ for all
  $j$ and $\lim_{j\to\infty} M_j/N_j=\lambda$ for some $0\leq \lambda
  \leq 1$. Then
\begin{equation}\label{s1}
\liminf_{j\to\infty} \frac 1{N_j} E_0(M_j,N_j) \geq \lambda \Engp(\lambda a)\, .
\end{equation}
\end{prop}
Note that (\ref{s1}) does not depend on $Z$. 

It is now useful to note that the energy $\Engp(a)$ is concave in $a$
(as an infimum over affine functions) and thus satisfies 
\begin{equation} \label{concave}
 \Engp(\lambda a) \geq (1-\lambda ) \Engp(0) + 
\lambda \Engp(a) \geq \lambda \Engp(a)\, .
\end{equation}
The last inequality in (\ref{concave}) follows from $\Engp(0) >0$.

The sequence $\M(N)$ defined above will have
a subsequence such that $\M(N_j)/N_j\to \lambda$ as $j\to \infty$ for
some $0\leq \lambda\leq 1$.  If we combine (\ref{MN})--(\ref{concave})
with $Z= 2\Engp(a) $ we find for this sequence that
\begin{equation}\label{lowerbound}
\limi_{j\to\infty}  E_0(N_j,N_j)/N_j \geq  \lambda^2 \Engp(a) +2(1-\lambda)\Engp(a)
=[ 1 + (1-\lambda)^2 ] \Engp(a) \geq \Engp(a) \, ,
\end{equation}
which proves (\ref{thmeq}) for this sequence $N_j$. (Here and in the following, we denote
$\liminf$ by $\limi$ for short.) Together with the upper bound
(\ref{upperbound}) we also conclude from (\ref{lowerbound}) that
$\lambda=1$. That is, for $Z\geq 2 \Engp(a)$ the sequence $\M(N)/N$
has only $1$ as a limit point, and hence (\ref{lowerbound}) holds for
the full sequence $N= 1,\, 2, \, 3\, ...$.

Our goal in the rest of this section is to prove
Proposition~\ref{mainlemma}, which then proves (\ref{thmeq}), as just
explained.

\subsection*{STEP 2. The Generalized Dyson Lemma.} 
 
To get a lower bound on $E_0(M,N)$, we start by deriving a lower bound
on the Hamiltonian $H_{\!_{M,N}}$, using Corollary~1 in \cite{LSS}. This
corollary, which is a generalization of Lemma 1 in \cite{LY1}
which, in turn, stems from Lemma 1 in Dyson's paper \cite{dyson}, asserts the
following. (Note that the range of the potential $v_{\!_N}$ is $R_0/N$, and
its scattering length is $a/N$. We use the ``hat'' $\widehat{\phantom
  X}$ to denote Fourier transform.)
\begin{lem}\label{dyson1}
  Let $R>R_0/N$. Let $\chi(\p)$ be a radial function such that $0\leq
  \chi (\p) \leq 1$ and such that $h(\x)\equiv \widehat{(1-\chi)}(\x)$
  is bounded and integrable (which implies that $\chi(\p) \to 1$ as
  $|p| \to \infty$).  Let
\begin{equation}\label{deffr} 
f_R(\x)= \sup_{|\y|\leq R} | h(\x-\y) - h(\x) |\,, 
\end{equation}
and
\begin{equation}\label{defwr} 
w_R(\x)= \frac{2}{\pi^2} f_R(\x) \int_{\R^3}
  f_R(\y)\,d\y\,.  
\end{equation} 
Let $U_R(\x)$ be any positive, radial function that vanishes outside
the annulus $R_0/N \leq |\x|\leq R$, with $\int_{\R^3} U_R(\x)\,d\x =
4\pi$.  Let $\eps>0$. If $\y_1,\dots,\y_n$ denote $n$ fixed points in
$\R^3$, with $|\y_i-\y_j|\geq 2R$ for all $i\neq j$, then we have the
operator inequality on $L^2(\R^3)$
\begin{equation}\label{dyson2}
-\nabla \chi(\p)^2 \nabla + \half \sum_{i=1}^n v_{\!_N}(\x-\y_i) \geq  
\sum_{i=1}^n 
\left( (1-\eps) \frac aN U_R(\x-\y_i) - \frac a{N\eps} w_R(\x-\y_i) \right)\,.
\end{equation}
\end{lem}
The sums in (\ref{dyson2}) are multiplication operators, i.e., they are
just functions of $x$. The operator $-\nabla \chi(\p)^2 \nabla$ is
just the positive multiplication by $p^2 \chi(\p)^2$ in Fourier space.
The original Lemma 1 in \cite{LY1} has $ \chi(\p) \equiv 1$ and
$h = w_R = f= \eps =0$.

\medskip
{\it Clarification:\/} What Lemma \ref{dyson1} really says is that we
can replace the unpleasant interaction potential $v_{\!_N}$ (which
possibly contains an infinite hard core) by a small, smooth, but
longer ranged potential whose main part, $U_R$, is positive. There
are two prices that have to be paid for this luxury. One is to forego
a piece of the positive kinetic energy, $-\nabla \chi(\p)^2
\nabla$. The second is that the potential is really only a `nearest
neighbor' potential. That is to say, the particle at $\x$ is allowed
to interact with only one other particle at a time.  This is seen from
the requirement that the interaction $U_R$ has range $R$, but the
other particles must be separated by a distance $2R$. In order to
utilize the coherent state inequalities later on in Step~3 we have to
extend our $U_R$ to an ordinary two-body potential, i.e., we have to
be able to drop the $2R$ separation requirement. To do so will require
an estimation of the amplitude (in the exact, original ground state
wave function) of finding three or more particles within a distance
$2R$ of each other. Clearly, this amplitude is small, but we find that
we have to resort to path integrals (or, more precisely, the Trotter
product formula) to estimate it. This will be done in Step~4 below.

As an immediate corollary of Lemma \ref{dyson1} we can omit the
condition $|y_i-y_j| \geq 2R$ and replace (\ref{dyson2})  by
\begin{align}\notag \label{dyson3}
&-\nabla \chi(\p)^2 \nabla + \half \sum_{i=1}^n v_{\!_N}(\x-\y_i) \\ &\geq  
\sum_{i=1}^n 
\left[(1-\eps) \frac aN U_R(\x-\y_i) - \frac a{N\eps} w_R(\x-\y_i) \right]
\prod_{k\neq i} \theta
  (|\y_k -\y_i|-2R)
\end{align}
for any set of points $y_j\in\R^3$. Here, $\theta$ denotes the
Heaviside step function, given by $\theta (t) =1$ if $t\geq 0$ and
$\theta (t) =0 $ if $t <0$.  That is to say, if there are only $n' <n$
of the $y_i$ that are a distance $\geq 2R$ from all the other $y_k$
then we simply apply (\ref{dyson2}) to these $n'$ coordinates. The
right side of (\ref{dyson3}) does not contain the other values of $i$
because the $\prod \theta$ factor vanishes for those. The left side
does contain these unwanted $y_i$ but, since $v_{\!_N}$ is
non-negative, this does no harm to the inequality (\ref{dyson3}).

We apply  (\ref{dyson3}) to each  particle, considering the other
$M-1$ particles as fixed, and obtain
\begin{multline}\label{ham2}
  H_{\!_{M,N}} \geq \sum_{i=1}^M \left( -\nabla_i \big(1-\chi(p_i)^2\big)
    \nabla_i + 2p_i\cdot \vecA(x_i)  +\vecA(\x_i)^2  + V(\x_i) \right)  \\ +
  \sum_{i=1}^M \sum_{j\neq i} \left[ (1-\eps) aN^{-1} U_R(\x_i-\x_j) -
    a (N\eps)^{-1} w_R(\x_i-\x_j) \right]\prod_{k\neq i,j} \theta
  (|\x_k-\x_j|-2R)\,.
\end{multline}
For the negative part of the interaction (containing $w_R$), we can
simply use $\prod \theta \leq 1 $ for a lower bound. 
For the positive part (containing $U_R$), we will use the fact that
\begin{equation}\label{3bod}
\prod_{k \neq i,j} \theta (|\x_k-\x_j|-2R) \geq 1 - \sum_{k \neq i,j} 
 \theta (2R-|\x_k-\x_j|)\, ,
\end{equation}
which follows from the simple inequality $\prod_j (1-s_j) \geq 1-
\sum_j s_j$ when $0\leq s_j\leq 1$ for all $j$.  

We now use (\ref{ham2}) and (\ref{3bod}) in the following way. 
 We begin by defining a new $M$-particle Hamiltonian, $\K$, by 
\begin{equation}\label{defK}
\K=  \sum_{i=1}^M \K_0^{(i)} + \sum_{1\leq i<j\leq M} 2(1-\eps) \,
aN^{-1} U_R(\x_i-\x_j) \,,
\end{equation}
where $\K_0$ is a one-body Hamiltonian to be described next.  If
$\K_0$ were simply $(-i\nabla+A)^2+V$ then (\ref{defK}) would be the
conventional Hamiltonian with two-body interaction $2U_R$. (The factor 2
arises because each pair $i,j$ appears twice in (\ref{ham2}).)

Unfortunately, $\K_0$ has to be a little more complicated because we 
used up part of the kinetic energy in replacing $v_{\!_N}$ by $U_R$ via Lemma 
\ref{dyson1}.  Pick some $\eta>0$, and let
\begin{equation}\label{k0}
\K_0=  -\nabla \big(1-\chi(\p)^2\big) \nabla -2 \eta \Delta + 
2p\cdot \vecA(x)  +\vecA(\x)^2 + V(\x)+ \eta |\x|^4 - \kappa(\eta)\,.
\end{equation}
The constant $\kappa(\eta)$ is chosen so that $K_0 >0$. It is a matter of
convenience to include it in the definition of $K_0$. It is defined by
\begin{equation}\label{24a}
\kappa(\eta)= \infspec \left[ -\eta\Delta + 2p\cdot \vecA(x) + 
\eta |\x|^4 \right]\,.
\end{equation}
The reason for adding the terms $-2\eta\Delta$ and $\eta |\x|^4$ to
$K_0$ is to ensure that $K_0$ is bounded from below and has compact
resolvent, and so that $\kappa(\eta)$ is finite. (Note: the exponent
$4$ in $|\x|^4$ could be replaced by any exponent $>2$ for our
purposes. This is due to the fact that we have a vector potential
$A(x)$ in mind that is bounded by $(\const)|x|$, as in the case of pure
rotation. If this is not so (because an external magnetic field has
been added) some polynomial of higher order than $|\x|^4$ could be
needed, but our analysis would continue to go through.)

Since there is a $2\eta \Delta $ in (\ref{k0}) and not just $\eta
\Delta $ we have that $K_0\geq -\eta \Delta+ V(x) \geq -\eta \Delta \geq 0 $,
since $V(x)\geq 0$ by assumption. This will be convenient later.

Let $\langle\,\cdot\,\rangle_\Bb$ denote the $M$-particle {\it
  bosonic\/} ground state expectation for the {\it original\/} Hamiltonian
$H_{\!_{M,N}}$. Actually, it is convenient to take the zero
temperature limit of the Gibbs state, which means that in case of a
ground state degeneracy of $H_{\!_{M,N}}$, we would take
$\langle\,\cdot\,\rangle_\Bb$ to be the uniform average over all
ground states. Then $ E_0(M,N) = \langle\, H_{\!_{M,N}} \,\rangle_\Bb$
and we have, therefore, using (\ref{ham2})--(\ref{24a}),
\begin{multline}\label{mainineq}
  E_0(M,N)\geq \infspec \K + M \kappa(\eta) - \eta M\big\langle
  |\x_1|^4\big\rangle_\Bb - 2 \eta M \big\langle
  -\Delta_1\big\rangle_\Bb \\ - \frac {M^2a}{N\eps} \big\langle
  w_R(\x_1-\x_2)\big\rangle_\Bb - \frac{a M^3}{N} \big\langle
  U_R(\x_1-\x_2) \theta(2R-|\x_2-\x_3|)\big\rangle_\Bb\,.
\end{multline}
(Note: We made use of the bosonic symmetry to replace $\sum_i \Delta_i$
by $M \Delta_1$, for example.)

The term $\langle -\Delta_1\rangle_\Bb$ can be bounded as follows. We
have $p^2\leq 2 (p+\vecA)^2 + 2 \vecA^2$, and hence, using positivity
of the interaction potential $v_{\!_N}$,
\begin{equation}\label{22a}
M \big\langle -\Delta_1\big\rangle_\Bb \leq 2 E_0(M,N) + \half {|\Omega|^2}  M 
\big\langle |\x_1|^2 \big\rangle_\Bb \,.
\end{equation}

To prove Proposition \ref{mainlemma} we have to bound the various terms in
(\ref{mainineq}) and (\ref{22a}), and that is what we do in the following steps.
The main term to bound is $\infspec \K $, the ground state energy of the 
`effective Hamiltonian' (\ref{defK}).

The momentum cutoff $\chi(p)$ in (\ref{k0}) will be chosen as follows. Let $\ell
(p)$ be an infinitely differentiable, spherically symmetric function
with $\ell(p)=0$ for $|p|\leq 1$, $\ell(p)=1$ for $|p|\geq 2$ and
$0\leq \ell(p)\leq 1$ in-between.
Then, for some adjustable parameter
$s$ to be determined later, we choose
\begin{equation}
\chi(p) = \ell(s p) \,.
\end{equation}
The potential $w_R(x)$  defined in (\ref{defwr}) is then a smooth and rapidly decreasing
function that depends only on the ratio $R/s$. It is easy to see that
\begin{equation}\label{wrint}
\int_{\R^3} w_R(\x)d\x \leq \const \frac {R^2}{s^2} 
\end{equation}
as long as $R\leq \const s$. We will, in fact, choose $R\ll s$.

Finally, we are still free to make a choice for the function $U_R(\x)$ in Lemma~\ref{dyson1}.
We choose it to be a `hat' function:
\begin{equation}
U_R(x) = \left\{ \begin{array}{ll} 6 R^{-3} & R\geq |x|\geq 2^{-1/3}R \\ 
0 & {\rm otherwise}\,,
\end{array}\right.
\end{equation}
assuming that $R \geq 2^{1/3} R_0/N$, a condition that will be amply
satisfied by our choice $N^{-1/3}\gg R\gg N^{-2/3}$ later on. We
remark that the exact form of $U_R(\x)$ is unimportant in what is to
come. We will need only the properties that $\int U_R(\x)d\x = 4\pi$
and that $\|U_R\|_\infty \leq \const R^{-3}$ for $R\gg R_0/N$.

\subsection*{STEP 3. Coherent State Method for the Ground State.}
We begin our analysis of (\ref{mainineq}) by bounding the main term, 
$\infspec \K $. This will be done with the aid of coherent states,
exploiting ideas in \cite{LSYjust}, and is, perhaps, the most methodologically novel
part of our work.

The one-body operator $K_0$ has purely discrete spectrum and can be
written in terms of its eigenvalues $e_j$ and orthonormal
eigenfunctions $|\varphi_j\rangle$ as $K_0= \sum_{j\geq 1} e_j
|\varphi_j\rangle\langle \varphi_j|$.  Recall that $K_0\geq
-\eta\Delta +V(x)\geq -\eta \Delta \geq 0$, so $e_j> 0$. We assume the
sequence $e_j$ to be ordered, i.e., $e_{j+1}\geq e_j$ for all $j$.
For simplicity, we introduce the notation
\begin{equation}
W(\x_1-\x_2)\equiv (1-\eps)a N^{-1} U_R(\x_1-\x_2)\,.
\end{equation}

The well known second quantization formalism involves the operators
$\ad{j}$ and $\an{j}$ which are the creation and annihilation
operators of a particle in the state $|\varphi_j\rangle$. They satisfy
the usual canonical commutation relations $[\an{i}, \ad{j}]
=\delta_{ij}$, etc.  The second quantized version of (\ref{defK}) is
\begin{equation}
\widehat \ham = \sum_{j\geq 1} e_j \ad{j} \an{j} + \sum_{ijkl} 
\ad{i}\ad{j}\an{k}\an{l} W_{ijkl}\,,
\end{equation}
where $W_{ijkl}= \langle \varphi_i\otimes \varphi_j| W |
\varphi_k\otimes \varphi_l\rangle$. The operator $\widehat \ham $ acts on the
bosonic Fock space, $\F$, consisting of a direct sum over all particle
number sectors.  We are interested in a lower bound to the ground
state energy of $\widehat \ham$ in the sector of particle number $M$.
Hence we can add a term $(\sum_j \ad{j}\an{j} - M)^2$ to $\widehat
\ham $ without changing this energy. We can then look for a lower
bound {\it irrespective of particle number}. I.e., for any $C\geq 0$, we have that $\infspec
\widehat \ham $ for $M$ particles is $\geq \infspec \ham$ on the {\it
  full\/} Fock space, where
\begin{equation}\label{choice}
 \ham \equiv  \sum_{j\geq 1} e_j \ad{j} \an{j} + \sum_{ijkl} \ad{i}
\ad{j}\an{k}\an{l} W_{ijkl}+ \frac {C}{M} \Big( \sum_{j\geq 1} \ad{j}
\an{j} - M\Big)^2 \,.
\end{equation}
The choice of $C$ will be  made later.

The Fock space $\F$ can be thought of as the tensor product of the
Fock spaces generated by each mode $\varphi_j$.  We choose some integer $J\gg
1$ (to be determined later) and split the Fock space into two parts,
namely $\F = \F^<\otimes \F^>$, where $ \F^<$ is the tensor product of
the Fock spaces generated by all the modes $\varphi_j$ with $j\leq J$
and where $ \F^>$ is that generated by all the other modes.

Next, we introduce coherent states \cite{Klauder} for all the modes
$j\leq J$. (By coherent states we mean ordinary canonical
Schr\"odinger, Bargmann, Glauber, coherent states.)  The modes with $j
>J$ will not be omitted, but they will be treated differently from the
$j\leq J$ modes.  Let $\vecz =(z_1,\dots,z_J)$ denote a vector in
$\C^J$. Let also $\Pi(\vecz)$ denote the projection onto the
coherent state $|z_1\otimes\cdots\otimes z_J\rangle \in \F^<$.  The
symbol $|z_1\otimes\cdots\otimes z_J\rangle$ is shorthand for
$|z_1\rangle
\otimes |z_2 \rangle \otimes\cdots\otimes |z_J\rangle$, and $|z_j
\rangle $ denotes the coherent state for the $j^{th}$ mode given by
$|z_j \rangle = \exp[-|z_j|^2/2+z_j \ad{j}]\, |\mathrm{vacuum} \rangle $.

The Hamiltonian $\ham$ in (\ref{choice}) can now be written as 
\begin{equation}\label{cho}
 \ham= \int d\vecz\, \Pi(\vecz)\otimes U(\vecz)\,,
\end{equation}
where $U(\vecz)$ is an operator acting on $\F^>$. The operator
$U(\vecz)$ depends on $\vecz$ since it is also an {\it upper symbol\/} for the
modes $j\leq J$. The integration measure is  $d\vecz =\pi^{-J}\prod_{j\leq J} dx_j
dy_j$ with $z_j=x_j+iy_j$.  This is discussed in \cite{LSYjust, Klauder}.
As an example, the upper symbol for $\ad{j}$ is ${\bar z}_j$ and for
$\an{j}$ it is $z_j$, but for $\ad{j} \an{j}$ it is $|z_j|^2
-1$. Thus, to a term such as $\ad{i}
\ad{j} \an{k } \an{l}$ with $i,j \leq J $ and $k,l >J$ would
correspond the upper symbol operator ${\bar z}_i\, {\bar z}_j\, \an{k
} \an{l}$.

It is easier to compute the {\it lower symbol\/} (which is denoted by
$u(\vecz)$) than the upper symbol $U(\vecz)$. It is obtained simply by
replacing $\ad{j}$ by ${\bar z}_j$ and $\an{j}$ by $z_j$ in all
(normal-ordered) polynomials, even in higher polynomials such as
$\ad{j}\an{j}$ or $\ad{j}\ad{j} \an{j} \an{j}$ . An equivalent
definition of the lower symbol of any polynomial $\mathcal{P}$ in the
$\an{j}$'s and $\ad{j}$'s (normal-ordered or not) is the expectation
value $u(\vecz) = \langle z_1\otimes\cdots\otimes z_J | \mathcal{P} |
z_1\otimes\cdots \otimes z_J\rangle$.  In the case considered here, $u(\vecz)= \langle
z_1\otimes\cdots\otimes z_J | \ham | z_1\otimes\cdots \otimes
z_J\rangle$.

The lower symbol is useful because the upper symbol can
conveniently be obtained from it as \cite{Klauder} 
\begin{equation}\label{symb}
U(\vecz) = e^{-\partial_\vecz \partial_{\bar\vecz}} u(\vecz) = u(\vecz) 
- \partial_{\vecz}\partial_{\bar\vecz}u(\vecz) + \half \left( 
\partial_{\vecz}\partial_{\bar\vecz}\right)^2 u(\vecz)\,,
\end{equation}
where $\partial_\vecz \partial_{\bar\vecz} = \sum_{j\leq J}
\partial_{z_{j}}\partial_{{\bar z}_j}$.  (In the general case there
would be higher order derivatives on the right side of
(\ref{symb}), but not in our case since $u(\vecz)$ is a
polynomial of order four.) Note that (\ref{cho}) implies that
\begin{equation}
\infspec \ham \geq \inf_{\vecz} \left( \infspec U(\vecz)\right) \,, 
\end{equation}
since $\int d\vecz\, \Pi(\vecz) = \id_{\F^<}$ and $ \Pi(\vecz)\otimes U(\vecz) \geq
\left[\infspec U(\vecz)\right] \Pi(\vecz)$. 

Our goal in the rest of this subsection is to derive a lower bound to
$\infspec U(\vecz)$ for a fixed $\vecz$. The reader might wonder why
we use coherent states only for modes $j\leq J$ and not for all modes.
The reason is that the upper symbol for the operator $e_j
\ad{j}\an{j}$ is $e_j (|z_j|^2 -1)$, and the $-1 $ term is a term that
we do not want when minimizing for a fixed $\vecz$. We make an error
in the energy of the form $-\sum_{j\leq J} e_j$ and for this reason we cannot take
$J=\infty$. But we can, and will let $J\to \infty$ as $N \to \infty$.

\subsubsection*{3a. Lower Bound on the Lower Symbol $u(\vecz)$.}  
In order to derive a lower bound to $U(\vecz)$ and the bottom of its
spectrum, we start by deriving a lower bound to the lower symbol
$u(\vecz)$, which is the first term in (\ref{symb}). This symbol can
be conveniently expressed in terms of the function $\Phi_\vecz \in
L^2(\R^3)$, parametrized by $\vecz \in \C^J$, given by
\begin{equation}
\Phi_\vecz(\x) = \sum_{1\leq j\leq J} z_j \varphi_j(\x)\,.
\end{equation}
Note that $\|\Phi_\vecz\|^2_2 =\sum_{j\leq J} |z_j|^2$.

Denoting $T\equiv \sum_{k>J} e_k \ad{k}\an{k}$, we have 
\begin{equation}\label{oneb}
\left\langle z_1\otimes\cdots\otimes z_J \left| \mbox{$\sum_{j}$} e_j \ad{j} \an{j} 
  \right| z_1\otimes\cdots\otimes z_J\right\rangle
= \sum_{j\leq J} e_j |z_j|^2 + T=\langle \Phi_\vecz | K_0 | \Phi_\vecz \rangle + T \,.
\end{equation}
There is a mild abuse of notation here, which will continue for the
rest of this paper, and which we hope will not cause any confusion.
The operator $ \sum_{j} e_j \ad{j} \an{j}$ acts on $\F$ while the
vector $ | z_1\otimes\cdots\otimes z_J \rangle$ is in $\F^<$, so the
left side of (\ref{oneb}) defines an operator on $\F^>$ in an obvious
way (actually, it defines a quadratic form).  The right side must
also be an operator on $\F^>$, and it is so if the number $\langle
\Phi_\vecz | K_0 | \Phi_\vecz \rangle $ is regarded as a number times
the identity on $\F^>$.

Similarly, with $N^>\equiv \sum_{j>J}\ad{j}\an{j}$ denoting the number of
particles in the modes $>J$,
\begin{align}\label{twob}
\Big\langle z_1\otimes\cdots\otimes z_J \Big| \Big( \mbox{$\sum_{j}$} 
\ad{j}\an{j} - M\Big)^2 \Big|z_1\otimes\cdots\otimes z_J\Big\rangle 
&= 
\left( N^> + \|\Phi_\vecz\|_2^2 - M \right)^2 + \|\Phi_\vecz\|_2^2 \notag \\ 
&\geq 
\left( \|\Phi_\vecz\|_2^2  - M \right)^2  - 2 e_J^{-1} M T \,.
\end{align}
Here, we used the normal ordering $[\sum_{j\leq J}\ad{j}\an{j}]^2 =
\sum_{i\leq J}\sum_{j\leq J}\ad{i} \ad{j}\an{i} \an{j} + \sum_{j\leq
  J}\ad{j}\an{j}$, followed by the elementary bound $N^>\leq T/e_J$.

The interaction part of $u(\vecz)$ is obtained by replacing 
$\an{j}$ by $z_j$ and $\ad{j} $ by ${\bar z}_j$ when $j\leq J$. 
We will now derive a lower bound on this term. It is convenient to
introduce the  notation
\begin{equation}
 I(\Phi_\vecz) =  \int d\x d\y  \,
|\Phi_\vecz(\x)|^2 |\Phi_\vecz(\y)|^2 W(\x-\y) \,.
\end{equation}
Since $W\geq 0$, it is possible to neglect the interaction between
modes $>J$ for a lower bound.  More precisely, let $P= \sum_{1\leq i\leq J}
|\varphi_i\rangle\langle \varphi_i|$ and $Q=1-P$.  The two-body
operator $W(\x-\y)$ is then bounded from below by
\begin{align}\label{wbound}
  W &= ((P+Q)\otimes (P+Q)) W ((P+Q)\otimes (P+Q)) \notag \\
&\geq (P\otimes P) W (P\otimes P) + (P\otimes P) W \left( P\otimes Q + 
Q\otimes P + Q\otimes Q\right) \notag \\
&\quad + \left( P\otimes Q + Q\otimes P + Q\otimes Q\right)W (P\otimes P)  \, , 
\end{align}
since the missing term on the right side of (\ref{wbound}) is
$(Q\otimes Q + P\otimes Q+ Q\otimes P ) W (Q\otimes Q +
 P\otimes Q+ Q\otimes P ) \geq 0$.
We thus have that 
\begin{multline}\label{low11}
\left\langle z_1\otimes\cdots\otimes z_J \left| \mbox{$\sum_{ijkl}$}  
\ad{i}\ad{j}\an{k}\an{l} W_{ijkl}  \right|z_1\otimes\cdots\otimes z_J\right\rangle \\
 \geq 
I(\Phi_\vecz)  + \sum_{kl>J}\langle \Phi_\vecz\otimes \Phi_\vecz| W|
  \varphi_k\otimes \varphi_l\rangle \an{k}\an{l} + \sum_{kl> J}
  \langle\varphi_k\otimes \varphi_l| W| \Phi_\vecz\otimes
  \Phi_\vecz\rangle \ad{k}\ad{l} \\ + 2 \sum_{k>J} \langle
  \Phi_\vecz\otimes \Phi_\vecz| W| \Phi_\vecz\otimes \varphi_k\rangle
  \an{k} + 2 \sum_{k>J} \langle \Phi_\vecz\otimes \varphi_k| W|
  \Phi_\vecz\otimes \Phi_\vecz\rangle\ad{k}\,.
\end{multline}
Here we used that $W$ is symmetric, implying that in the last line we
could replace $|\Phi_\vecz\otimes \varphi_k + \varphi_k \otimes
\Phi_\vecz\rangle $ by $2 |\Phi_\vecz\otimes\varphi_k\rangle$.

We seek a lower bound to the last two expressions in
(\ref{low11}). Note that, for a general operator $A$, $|A+A^\dagger|^2
= A^2+A^{\dagger 2} +A A^\dagger + A^\dagger A \leq 2 A^\dagger A + 2
A A^\dagger$ by Schwarz's inequality, and so
\begin{equation}
(A+A^\dagger)^2 \leq 4  |A|^2 + 2 [A,A^\dagger]\, .
\end{equation}
 We apply this to the second line in
(\ref{low11}), with $A= \sum_{kl>J} c_{kl} \an{k}\an{l}$ and
$c_{kl}=\langle \Phi_\vecz\otimes \Phi_\vecz| W| \varphi_k\otimes
\varphi_l\rangle$. The commutator is
\begin{equation}\label{commu}
[A,A^\dagger] = 4 \sum_{klm>J} \ad{k} \an{l} \overline{c_{km}} c_{lm} + 
2 \sum_{kl>J} |c_{kl}|^2 \,.
\end{equation}
The last term in (\ref{commu}) is bounded by
\begin{equation}
\sum_{kl> J} |c_{kl}|^2 \leq \sum_{kl\geq 1} |c_{kl}|^2 = \langle 
\Phi_\vecz\otimes \Phi_\vecz| W^2 |\Phi_\vecz\otimes \Phi_\vecz\rangle 
\leq  \|W\|_\infty I(\Phi_\vecz) \,.
\end{equation}
The first term on the right side of (\ref{commu}) can be bounded as  
\begin{equation}\label{imp}
\sum_{klm>J} \ad{k} \an{l} \overline{c_{km}} c_{lm} \leq  \sum_{m\geq 1} \sum_{kl>J} 
\ad{k} \an{l} \overline{c_{km}} c_{lm}  \leq  \frac {4}{27 \pi^4} \eta^{-1} 
\|W\|_1^2 \|\nabla\Phi_\vecz\|_2^4 \, T \,.
\end{equation}
This can be seen as follows. The integral kernel $\sigma$ of the one-particle
operator defined by the matrix $\sum_{m\geq 1} \overline{c_{km}} c_{lm}$ is
given by
\begin{equation}
\sigma(x,x') =\int dy |\Phi_\vecz(y)|^2 W(x-y) W(x'-y) 
\overline{\Phi_\vecz(x)} \Phi_\vecz(x')\,.
\end{equation}
Using Young's and Schwarz's inequalities, we have, for any function $f$ on $\R^3$,
\begin{align}
 \int dx\, dx' \overline{f(x)} f(x') \sigma(x,x') 
&\leq \int dx\, dx'\, dy |\Phi_\vecz(y)|^2 W(x-y) W(x'-y)
  |\Phi_\vecz(x) f(x)|^2 \notag\\
&\leq \|W\|_1^2 \|\Phi_\vecz\|_6^{2} \|
  \Phi_\vecz f\|_{3}^{2} \  \leq \  \|W\|_1^2 \|\Phi_\vecz\|_6^{4}
  \|f\|_6^{2}\,.
\end{align}
Hence (\ref{imp}) follows by applying the Sobolev inequality
$\|f\|_6^2 \leq (4/3) (2\pi^2)^{-2/3} \|\nabla f\|_2^2$ both to $f$
and to $\Phi_\vecz$, and using the fact that $-\Delta \leq \eta^{-1} K_0$.

To get an upper bound on $|A|^2$ we use Schwarz's
inequality again to obtain
\begin{equation}
|A|^2 \leq \left(\sum_{kl>J} \frac {|c_{kl}|^2}{e_k e_l}\right)
\left( \sum_{mn>J} e_m e_n \ad{m}\ad{n}\an{m}\an{n}\right)
\end{equation}
for any sequence of positive numbers $e_j$. We choose $e_j$ to be the
eigenvalues of $K_0$, in which case
\begin{equation}
 \sum_{mn>J} e_m e_n \ad{m}\ad{n}\an{m}\an{n} \leq 
\left( \sum_{k>J} e_k \ad{k}\an{k} \right)^2 = T^2\,.
\end{equation}
Moreover,
\begin{equation}\label{ckl}
\sum_{kl>J} \frac {|c_{kl}|^2}{e_k e_l} = \left\langle \Phi_\vecz\otimes 
\Phi_\vecz \left| W \left( \frac Q {K_0} \otimes \frac Q{K_0}\right) W \right|  
\Phi_\vecz\otimes \Phi_\vecz \right\rangle \,.
\end{equation}
We have the following two operator inequalities; the first comes from
the fact that $K_0 \geq e_J$ on the range of the projector $Q$ and the
second comes from $K_0 \geq -\eta \Delta$:
\begin{equation}\label{triq}
\frac Q{K_0} \leq \frac 2{K_0 + e_J} \leq \frac 2{-\eta \Delta + e_J}\,.
\end{equation}
Denoting the integral kernel of $(-\Delta +\mu)^{-1}$ by 
\begin{equation}
k_\mu(x-x') = \frac 1{4\pi} \frac{ e^{-\sqrt{\mu}|\x-\x'|}}{|\x-\x'|}\,,
\end{equation} 
we see that (\ref{ckl}) is bounded above by
\begin{align}
 &\frac 4{ \eta^{2}} \int d\x d\y d\x' d\y'\, \overline{\Phi_\vecz(\x)}
  \overline{\Phi_\vecz(\y)} W(\x-\y) k_{e_J/\eta}(x-x')k_{e_J/\eta}
(\y-\y')\Phi_\vecz(\x')\Phi_\vecz(\y') W(\x'-\y') \notag \\ 
  &\leq \frac 4{ \eta^{2}}\int d\x d\y d\x' d\y'\, |\Phi_\vecz(\x)|^2
  |\Phi_\vecz(\y)|^2 W(\x-\y)k_{e_J/\eta}(x-x')k_{e_J/\eta}(\y-\y') 
W(\x'-\y') \notag  \\ 
&\leq \frac 1{2\pi \eta^{3/2}} \frac {\|W\|_1}
{\sqrt{e_J}} I(\Phi_\vecz)\,.
\end{align}
Here, we used Young's inequality for the $(x',y')$ integration, as
well as the fact that $\|k_\mu\|_2^2=(8\pi\sqrt{\mu})^{-1}$.  

By putting all this together, we have that
\begin{equation}\label{sqrttr}
(A+A^\dagger)^2 \leq 4  \|W\|_\infty I(\Phi_\vecz) + 
\frac {32}{27 \pi^4} \eta^{-1} \|W\|_1^2 \|\nabla\Phi_\vecz\|_2^4 \, 
T+ \frac 4{2\pi \eta^{3/2}} \frac {\|W\|_1}{\sqrt{e_J}} I(\Phi_\vecz)\, 
T^2  \, .
\end{equation}
Since the square root preserves operator
monotonicity, we can take the square root on both sides of
(\ref{sqrttr}). By the triangle inequality, we can take the sum of the
square roots of each term on the right side. Finally, applying the
Schwarz inequality to the first and third term, we conclude that, for any $\delta>0$,
\begin{align}
|A+A^\dagger| \leq \delta I(\Phi_\vecz) &+ \frac 1\delta \|W\|_\infty + 
\frac 4{\pi^2} \sqrt{\frac {2}{27}} \eta^{-1/2} \|W\|_1  \|\nabla 
\Phi_\vecz\|_2^2 \sqrt{T}    \notag \\
&+ e_J^{-1/4} \left( \|W\|_1 I(\Phi_\vecz) 
+ \big(2\pi\eta^{3/2}\big)^{-1}\right)\, T \, .
\end{align}

We now proceed similarly with the last term in (\ref{low11}) which is
linear in $\an{k}$ and $\ad{k}$. Denoting $c_k=\langle
\Phi_\vecz\otimes \Phi_\vecz| W| \Phi_\vecz\otimes \varphi_k\rangle$,
we have that
\begin{equation}\label{22}
\left( \sum_{k>J} \left( c_k \an{k} + \overline{c_k} \ad{k}\right) \right)^2
\leq 4  \left(\sum_{k>J} \frac {|c_k|^2}{e_k} \right) 
\left( \sum_{k>J} e_k \ad{k}\an{k}\right) +2 \sum_{k>J} |c_k|^2 \,.
\end{equation}
Using  H\"older's and Sobolev's inequality,
\begin{align}\label{58}
  \sum_{k\geq 1} |c_k|^2 &= \int d\x d\y d\z \, |\Phi_\vecz(\x)|^2
  |\Phi_\vecz(\y)|^2 |\Phi_\vecz(\z)|^2 W(\x-\y) W(\x-\z) \notag \\ 
&\leq
  \|W\|_{3/2} \|\Phi_\vecz\|_6^2 I(\Phi_\vecz) \ \leq \ \frac 4{3 (2\pi^2)^{2/3}} 
  \|W\|_{\infty}^{1/3}\|W\|_{1}^{2/3} \|\nabla\Phi_\vecz\|_2^2 I(\Phi_\vecz)
  \ .
\end{align}
Moreover, using (\ref{triq}) again, together with
Young's and Sobolev's inequalities, as well as the
fact that the $3/2$ norm of $k_\mu$ is given by
$2^{-1/3}\mu^{-1/2}/3$, we find that
\begin{align}\label{24}
  \sum_{k>J} \frac {|c_k|^2}{e_k}  &\leq \frac 2\eta \int d\x d\y d\x'
  d\y' \, \overline{\Phi_\vecz(\x)} |\Phi_\vecz(\x')|^2 W(\x-\x')
  k_{e_J/\eta}(\x-\y)
  \Phi_\vecz(\y)|\Phi_\vecz(\y')|^2 W(\y-\y') \notag \\
&\leq \frac 2\eta \int
  d\x d\y d\x' d\y' \, |\Phi_\vecz(\x)|^2 |\Phi_\vecz(\x')|^2
  W(\x-\x') k_{e_J/\eta}(\x-\y)
  |\Phi_\vecz(\y')|^2 W(\y-\y') \notag \\ 
&\leq  \frac 4{9 \pi^{4/3}} \eta^{-1/2} 
\frac { \|W\|_1}{
    \sqrt{e_J}} \| \nabla \Phi_\vecz\|_2^2 I(\Phi_\vecz) \,.
\end{align}
This implies that   
\begin{align}
  &\left( \sum_{k>J} \left( c_k \an{k} + \overline{c_k} \ad{k}\right) \right)^2  
\notag \\
  &\leq \frac 8{3 (2\pi^2)^{2/3}} |W\|_{\infty}^{1/3}\|W\|_{1}^{2/3}\|
  \nabla \Phi_\vecz\|_2^2 I(\Phi_\vecz)+ \frac {16}{9 \pi^{4/3}}
  \eta^{-1/2}
  \frac { \|W\|_1}{\sqrt{e_J}}  \|\nabla \Phi_\vecz\|_2^2 I(\Phi_\vecz)\, T   \label{sqr} \\
  &\leq \left(\sqrt{\frac 2{3}}\frac 1{(2\pi^2)^{1/3}}
    \|W\|_\infty^{1/6} \|W\|_1^{1/3} + \frac 2{3\pi^{2/3}} \eta^{-1/4}
    \|W\|_1^{1/2} e_J^{-1/4}\, \sqrt{T} \right)^2 \bigg( \|\nabla
  \Phi_\vecz\|_2^2 + I(\Phi_\vecz)\biggr)^2 \,, \notag
\end{align}
again using the triangle and the Schwarz inequality. As mentioned
above, operator monotonicity is preserved by the square root, and
hence we can take the square root on both sides of Eq. (\ref{sqr}).

This completes the lower bound on the lower symbol $u(\vecz)$. For the
convenience of the reader, we repeat the bound just derived:
\begin{align}\label{lbu}
u(\vecz) \geq &\langle \Phi_\vecz|K_0|\Phi_\vecz\rangle + I(\Phi_\vecz)
\left(1-\delta - e_J^{-1/4} \|W\|_1 T\right) + \frac CM 
\left( \|\Phi_\vecz\|_2^2 -M\right)^2  \\ 
&- \left( \|\nabla \Phi_\vecz\|_2^2 + I(\Phi_\vecz)\right)
\left(2\sqrt{\frac 2{3}}\frac 1{(2\pi^2)^{1/3}}  \|W\|_\infty^{1/6} 
\|W\|_1^{1/3} + \frac 4{3\pi^{2/3}} \eta^{-1/4} \|W\|_1^{1/2} e_J^{-1/4}\, 
\sqrt{T} \right) \notag \\
&- \|\nabla \Phi_\vecz\|_2^2 \frac 4{\pi^2} 
\sqrt{\frac 2{27}} \eta^{-1/2} \|W\|_1 \sqrt{T} - \frac 1\delta 
\|W\|_\infty + T\left( 1-e_J^{-1/4} 
\big(2\pi^2\eta^{3/2}\big)^{-1} - \frac{2C}{e_J}\right)  \,.   \notag
\end{align}
We note that in the following we will choose $J$ large enough so that
the last term in (\ref{lbu}) is positive and thus can be neglected for
a lower bound. (Recall that $e_J\to\infty$ as $J\to\infty$.)


\subsubsection* {3b. Lower Bound on the Remaining Terms in $U(\vecz)$.} 
A lower bound on the first term on the right side of (\ref{symb}) is
given in (\ref{lbu}) and, therefore, to get a lower bound on the upper
symbol $U(\vecz)$, it remains to bound the last two terms on the
right side of (\ref{symb}). The very last term is positive, as will be
shown now, and can thus be neglected for a lower bound. Namely,
\begin{align}\label{seco}
  \half \left( \partial_{\vecz}\partial_{\bar\vecz}\right)^2 u(\vecz) &= \half
  \left( \partial_{\vecz}\partial_{\bar\vecz}\right)^2 \left(
    \langle \Phi_\vecz\otimes \Phi_\vecz| W | \Phi_\vecz\otimes
    \Phi_\vecz\rangle + \frac CM \|\Phi_\vecz\|_2^4 \right)\\ &= \half 
  \sum_{1\leq ij\leq J} \langle\varphi_i\otimes \varphi_j+
\varphi_j\otimes \varphi_i | W |\varphi_i\otimes \varphi_j+\varphi_j\otimes 
\varphi_i\rangle + \frac {C}M J(J+1) \geq 0\,. \notag
\end{align}

The remaining expression, $\partial_{\vecz}\partial_{\bar\vecz}
u(\vecz)$, consists of the following terms. First, from the one-body
part (\ref{oneb}) of the Hamiltonian we obtain a contribution
$\sum_{j\leq J} e_j$.  Second, from the term (\ref{twob}) (see also
(\ref{choice})) that was introduced in order to control the particle
number, we get
\begin{equation}
\frac CM \left\{ (2N^> -2M +1 +2\|\Phi_\vecz \|^2)J + 2\|\Phi_\vecz\|^2\right\}
\leq  \frac {2 C}M (J+1)\|\Phi_\vecz\|_2^2 + \frac { J C}M  
\left(2N^>+1\right) \,.
\end{equation}

Finally, the following three contributions are obtained from the
interaction part. From the part where all four indices are $\leq J$,
we have
\begin{align}\label{alst}
  \sum_{j\leq J} &\langle \Phi_\vecz\otimes\varphi_j + \varphi_j\otimes
  \Phi_\vecz| W | \Phi_\vecz \otimes\varphi_j + \varphi_j\otimes
  \Phi_\vecz\rangle \leq 4 \sum_{j\leq J} \langle \Phi_\vecz\otimes
  \varphi_j | W | \Phi_\vecz \otimes \varphi_j \rangle  \notag \\ 
&\leq 4
  \sum_{j\leq J} \|W\|_1 \, \|\Phi_\vecz\|_6^2 \|\varphi_j\|_3^2 \ \leq \ 
  (4/3)^{3/2} \frac 1{2\pi^2} \eta^{-1/2} \|W\|_1 \,
  \|\nabla\Phi_\vecz\|_2^2 \sum_{j\leq J} \sqrt{e_j} \,.
\end{align}
Here, we used the inequalities of Young, H\"older and Sobolev as well
as the facts that $-\Delta\leq \eta^{-1} K_0$ and $\langle \varphi_j
|K_0 | \varphi_j \rangle = e_j$ in the last step.  {F}rom the term
with 3 indices $\leq J$, we get
\begin{equation}\label{isb}
2 \sum_{j\leq J} \sum_{ k>J} \langle \Phi_\vecz\otimes \varphi_j + 
\varphi_j\otimes \Phi_\vecz | W | \varphi_j \otimes 
\varphi_k\rangle \an{k} +  \mathrm{adjoint} \,.
\end{equation}
Using (\ref{22}), this time with $e_k\equiv 1$, (\ref{isb}) is bounded above,
as an operator, by 
\begin{equation}\label{ere}
4 \sum_{j\leq J} \left[ \big(N^>+\half \big) \sum_{k>J} \big |  \langle 
\varphi_j \otimes \varphi_k | W |  \Phi_\vecz\otimes \varphi_j + 
\varphi_j\otimes \Phi_\vecz\rangle |^2 \right]^{1/2} \,.
\end{equation} 
Similarly to (\ref{58}), we can derive the bound
\begin{equation}
\sum_{k\geq 1} |  \langle \varphi_i \otimes \varphi_k | W |  
\Phi_\vecz\otimes \varphi_i + \varphi_i\otimes \Phi_\vecz\rangle |^2 
\leq 4 \|W\|_{3/2} \|W\|_1 \|\Phi_\vecz\|_6^2 \|\varphi_i\|_6^3\|\varphi_i\|_2  \,.
\end{equation}
Since $\|\varphi_i\|_2=1$ and $\|\varphi_i\|_6^2 \leq (4/3)
(2\pi^2)^{-2/3} \|\nabla \varphi_i\|_2^2 \leq (4/3) (2\pi^2)^{-2/3}
\eta^{-1} e_i$, this implies that
\begin{align}
  (\ref{ere}) &\leq 8 (4/3)^{5/4} \frac 1{(2\pi^2)^{5/6}} \|W\|_1^{5/6}
  \|W\|_\infty^{1/6} \|\nabla\Phi_\vecz\|_2 \eta^{-3/4} \sum_{i\leq J}
  e_i^{3/4} \, \sqrt{N^>+\half } \notag \\ &\leq 4 (4/3)^{5/4} \frac
  1{(2\pi^2)^{5/6}} \|W\|_1^{5/6} \|W\|_\infty^{1/6} \eta^{-3/4}
  \sum_{i\leq J} e_i^{3/4} \left( \|\nabla\Phi_\vecz\|_2^2+ N^>+\half
  \right) \,.
\end{align}
Here, Schwarz's inequality was used in the last step. 

The last term to estimate is the one coming from 2 indices $\leq J$, given by
\begin{align}\notag
&\sum_{j\leq J}\sum_{ k,l>J} \langle \varphi_j\otimes \varphi_k + 
\varphi_k\otimes\varphi_j| W | \varphi_j\otimes \varphi_l + 
\varphi_l\otimes\varphi_j\rangle \ad{k}\an{l} \\ &\leq (4/3)^{3/2} 
\frac 1{2\pi^2} \eta^{-3/2} \|W\|_1 \sum_{j\leq J} \sqrt{e_j} \, T \, . 
\end{align}
This inequality can be seen as follows. For any one-particle function $f$, 
\begin{align}
\langle \varphi_i\otimes f + f\otimes\varphi_i| W | \varphi_i\otimes f 
+ f\otimes\varphi_i\rangle  &\leq 4 \langle \varphi_i\otimes f | W | 
\varphi_i\otimes f \rangle \ \leq \ 4 \|W\|_1 \|f\|_6^2 \|\varphi_i\|_3^2 
\notag \\ 
&\leq (4/3)^{3/2} \frac 1{2\pi^2} \eta^{-1/2} \|W\|_1 \|\nabla f\|_2^2  
\sqrt{e_i} \,.
\end{align}
The last inequality is the same is in (\ref{alst}). The result now
follows using $-\Delta \leq \eta^{-1} K_0$.  

Altogether, we have thus shown that
\begin{align}\label{uno}
 \partial_{\vecz}\partial_{\bar\vecz} u(\vecz) \leq  \sum_{i\leq J} & e_i + 
\frac {2 C}M (J+1)\|\Phi_\vecz\|_2^2 + \frac {2 C J}M  
\left( N^>+\half\right) \notag \\ 
&+  (4/3)^{3/2} \frac 1{2\pi^2} \eta^{-1/2} \|W\|_1
\left( \|\nabla\Phi_\vecz\|_2^2 
+ \eta^{-1} T \right) \sum_{i\leq J} \sqrt{e_i} 
\\ &+ 4 (4/3)^{5/4} \frac 1{(2\pi^2)^{5/6}} \|W\|_1^{5/6} \|W\|_\infty^{1/6}
 \eta^{-3/4} \sum_{i\leq J} e_i^{3/4}  
\left(  \|\nabla\Phi_\vecz\|_2^2+ N^>+\half 
\right) \,.  \notag 
\end{align}
This finishes our lower bound on the upper symbol $U(\vecz)$. 
To summarize, we have shown the following operator lower bound to the operator
$U(\vecz)$:
\begin{equation}\label{Ubound}
U(\vecz) \geq  {\rm right\ side\ of\ } (\ref{lbu})\ - \  {\rm right\ side\ of\ } (\ref{uno})\ .
\end{equation}

\subsubsection*{3c. $c$-Number Bound on $T$.}
We are interested in the ground state energy of $U(\vecz)$ for a fixed
$\vecz \in \C^J$. Since $T$ and $N^>$ are the only operators appearing
in (\ref{lbu}) and (\ref{uno}), this quantity can be bounded from below
using (\ref{lbu}) and (\ref{uno}) if we can evaluate the expectation
values of $T$ and $N^>$ in the ground state (or one of the ground
states) of $U(\vecz)$.

Let $\langle \ \cdot\ \rangle_\vecz$ denote the expectation value in a
ground state of $U(\vecz)$. We can use two simple facts: i.)  Since
$\sqrt{T}$ enters (\ref{lbu}) negatively, we can use the concavity of
the square root to replace $\langle \sqrt{T} \rangle_\vecz$ by $\sqrt{
  \langle T\rangle_\vecz}$ for a lower bound.  \ ii.) Since $N^>$ appears positively
in (\ref{uno}), and hence negatively in (\ref{Ubound}), we can replace
it by the upper bound $N^> \leq T/e_J $.

For the purpose of bounding $\langle T \rangle_\vecz$ we can use a
lower bound to $U(\vecz)$ that is much simpler than (\ref{Ubound}).
This is obtained by totally neglecting both the interaction part and
the part controlling the particle number in $u(\vecz)$.  These give
positive contributions to $u(\vecz)$ (since $u(\vecz)$ is the
expectation value of $\ham$ in the coherent state). We have to be more
careful about $\partial_{\vecz}\partial_{\bar\vecz} u(\vecz)$,
however, because this contains some negative terms, as given in
(\ref{uno}).  (The annoying fact is that an upper symbol of a positive
operator need not be positive, although the lower symbol is always
positive.)

Proceeding in the manner just described we have that
\begin{equation}\label{swos}
U(\vecz) \geq 
\langle \Phi_\vecz | K_0 | \Phi_\vecz \rangle + T  -  
\partial_{\vecz}\partial_{\bar\vecz} u(\vecz) \,.
\end{equation}
Now let us estimate the various terms in $\partial_{\vecz}\partial_{\bar\vecz}
u(\vecz)$ in (\ref{uno}). We have $\eta \|\nabla\Phi_\vecz\|_2^2 
\leq\langle\Phi_\vecz | K_0 | \Phi_\vecz \rangle$. Also $\|
\Phi_\vecz\|_2^2 \leq \langle\Phi_\vecz | K_0 | \Phi_\vecz \rangle /
\infspec (-\eta\Delta+V(x))$. Moreover, $\|W\|_1 \leq 4\pi a/N$, and
$\|W\|_\infty \leq 6 a/(R^3 N)$. 

We will choose $R\gg N^{-2/3}$ below. Therefore, $\|W\|_\infty \ll N$.
The operator $N^>$ can be bounded in terms of $T$ as $N^>\leq T/e_J$.
Note also that $M=O(N)$ by assumption. Hence we see from (\ref{uno}) and (\ref{swos}) that,
for $N$ large enough (depending on the parameters $\eta$, $C$ and
$J$),
\begin{equation}\label{ut}
U(\vecz) \geq \half T - \const \,,
\end{equation}
where the constant depends only on  $\eta$, $C$ and $J$,
but not on $M$ or $N$.

The value of (\ref{ut}) is that it allows us to control the value of
$\langle T\rangle_\vecz$, and thereby control $\inf_\vecz \infspec
U(\vecz)$, which is our lower bound to the ground state energy of
$\ham$. There is some number $E$, independent of all parameters, such
that $\inf_\vecz \infspec U(\vecz) \leq ME/2 -\const$ because
$\inf_\vecz \infspec U(\vecz)$ is less than the known upper bound to
the ground state energy of $\ham$. Then we can, and will restrict our
attention to $\vecz$'s with $\langle T\rangle_\vecz \leq ME$ because
only those values of $\vecz$ are relevant for computing $\inf_\vecz
\infspec U(\vecz)$, as (\ref{ut}) shows.  Only the existence of $E$ and
not its value is important.

We conclude from (\ref{Ubound}) and the fact that $\langle
T\rangle_\vecz \leq M E$ for the $\vecz$ in question that
\begin{equation}\label{ephi1}
\inf_\vecz \infspec U(\vecz) \geq \inf_\Phi \E[\Phi] \,,
\end{equation}
where 
\begin{equation}\label{ephi2}
\E[\Phi] = \langle \Phi|K_0|\Phi\rangle + D_1\, I(\Phi) + \frac CM 
\left( \|\Phi\|_2^2 -M\right)^2 - D_2 \|\nabla\Phi\|_2^2 - D_3  - \frac {2 C}M 
(J+1)\|\Phi\|_2^2\,.
\end{equation}
The notation is the following:
\begin{align}\label{d1def}
D_1= 1&-\delta - e_J^{-1/4} \|W\|_1 M E - 2\sqrt{\frac 2{3}}\frac 1{(2\pi^2)^{1/3}}  
\|W\|_\infty^{1/6} \|W\|_1^{1/3}  \notag\\ 
&- \frac 4{3\pi^{2/3}} \eta^{-1/4} \|W\|_1^{1/2} 
e_J^{-1/4} M^{1/2} E^{1/2} \,,
\end{align}
\begin{align}\label{d2def}
D_2= 2&\sqrt{\frac 2{3}}\frac 1{(2\pi^2)^{1/3}}  \|W\|_\infty^{1/6} \|W\|_1^{1/3} 
+ \frac 4{3\pi^{2/3}} \eta^{-1/4} \|W\|_1^{1/2} e_J^{-1/4} M^{1/2} E^{1/2} \notag\\ 
&+ \frac 4{\pi^2} \sqrt{\frac 2{27}} \eta^{-1/2} \|W\|_1 M^{1/2} E^{1/2} 
+  (4/3)^{3/2} \frac 1{2\pi^2} \eta^{-1/2} \|W\|_1 \sum_{i\leq J} \sqrt{e_i} \notag \\ 
&+ 4 (4/3)^{5/4} \frac 1{(2\pi^2)^{5/6}} \|W\|_1^{5/6} \|W\|_\infty^{1/6}
 \eta^{-3/4} \sum_{i\leq J} e_i^{3/4} \,,
\end{align}
and
\begin{align}\label{d3def}\notag
D_3 = &  \sum_{i\leq J} e_i + \frac {2 C J}M  \left( ME/e_J+\half\right)+  (4/3)^{3/2} 
\frac 1{2\pi^2} \eta^{-3/2} \|W\|_1 ME \sum_{i\leq J} \sqrt{e_i}
\\ &+ 4 (4/3)^{5/4} \frac 1{(2\pi^2)^{5/6}} \|W\|_1^{5/6} \|W\|_\infty^{1/6}
 \eta^{-3/4} \sum_{i\leq J} e_i^{3/4} \left( ME e_J^{-1} +\half \right) 
+ \frac 1\delta \|W\|_\infty\,.
\end{align}
We have neglected the last term in (\ref{lbu}) containing $(
1-e_J^{-1/4} (2\pi^2\eta^{3/2})^{-1} - 2C/e_J)$, assuming $J$ to be
large enough to make this term positive. (Recall that $e_J\to \infty$
as $J\to \infty$.)

Our final result in this section, (\ref{ephi1})--(\ref{ephi2}), might
not appear to be useful at first sight, but the reader should note
that the first two terms in (\ref{ephi2}) are essentially the GP
energy expression. The term $\langle \Phi|K_0|\Phi\rangle $ is the
relevant (i.e., low momentum) part of the kinetic energy $\int |
(i\nabla-A)\Phi|^2$. The coefficient $D_1$ equals 1 to leading order
and $ I(\Phi)$ is essentially the GP quartic term $4 \pi a \int
|\Phi|^4$ (up to errors which will be controlled). Moreover, for $C$
large enough the term $C(\|\Phi\|_2^2-M)^2/M$ ensures that we have
the right particle number. For an appropriate choice of the parameters
$J$, $\eta$ and $R$ all other terms are of lower order as $N\to
\infty$, as we shall show.


\subsection*{STEP 4.  Bounds on Three-Particle Density.}  
So far we have bounded the main term in (\ref{mainineq}), namely
$\infspec \K$.  Of the various other terms in (\ref{mainineq}) that
have to be bounded, the one that is most intuitively negligible, but
which we find the hardest to control is the last term in
(\ref{mainineq}). To show that it is small we have to show that the
probability of finding three particles within a distance $2R$ of each
other (in a true ground state of $H_{\!_{M,N}}$) is small. This is
accomplished in this section.

We begin with a lemma about the possible size of the expectation value
of a function of the coordinates of three bosons.  Recall from Step 2 that
$\langle\ \cdot\ \rangle_\Bb $ denotes expectation value in the {\it
  bosonic}, zero-temperature state of the $M$-body Hamiltonian
$H_{\!_{M,N}}$.

\begin{lem}\label{3plemma}
  Let $\xi (\x_1,\x_2,\x_3)$ be any positive function of $\x_1$,
  $\x_2$ and $\x_3\in\R^3$. With $V=$ the one-body potential appearing
  in $H_{\!_{M,N}}$, we define the three-body, independent particle
  Hamiltonian $$ h= -\Delta_1-\Delta_2-\Delta_3 + V(\x_1)
  +V(\x_2)+V(\x_3)\ .  $$ Let $\alpha>0$ and let $e^{-\alpha
  h}(\x_1,\x_2,\x_3\ ; \
\y_1,\y_2,\y_3)$ be the `heat kernel' of $h$ at `inverse temperature'
$\alpha$. Finally, consider the modified integral kernel
\begin{equation}
e^{-\alpha h}(\x_1,\x_2,\x_3\ ; \ \y_1,\y_2,\y_3)
  \sqrt{\xi(\x_1,\x_2,\x_3) \xi( \y_1,\y_2,\y_3)}
\end{equation} 
and let $\Lambda$ denote its largest eigenvalue (i.e., its norm as a
map from $L^2(\R^9)$ to $L^2(\R^9)$).Then
\begin{equation}\label{lemeq2}
\big\langle \xi(\x_1,\x_2,\x_3) \big\rangle_\Bb \leq \Lambda 
\exp\{\alpha(E_0(M,N)-E_0(M-3,N))\}\, .
\end{equation}
\end{lem}

Note that for the $M$ and $N$ under consideration here, we have
$E_0(M,N)-E_0(M-3,N)\leq 3Z$, as explained in Step 1.  It is the
appearance of the peculiar difference $E_0(M,N)-E_0(M-3,N)$ in Lemma
\ref{3plemma} that led us to the discussion in Step 1. If the
three-body correlations could be bounded more expeditiously than is
done here, Step 1 could be simplified.

\begin{proof}
  We denote by $\Tr\,[\,\cdot\,]$ the trace over {\it all\/} of
  $L^2(\R^{3M})$, not just the bosonic states, and by $\pbos$ the
  projection onto the bosonic (i.e., symmetric) subspace. Note that
  $\exp\{-\beta H_{\!_{M,N}}\}$ is trace class for large enough
  $\beta$, by our assumption on the logarithmic increase of the
  potential $V(x)$. (This follows from the Feynman-Kac-It{\^o} formula,
  together with the results in the appendix.) Hence
\begin{equation}
\big\langle \xi \big\rangle_\Bb = \lim_{n\to\infty} \frac {\Tr\, [\xi 
e^{-\alpha n H_{\!_{M,N}}} \pbos]}{\Tr\, [e^{-\alpha n H_{\!_{M,N}}} \pbos]} \,,
\end{equation}
independently of $\alpha$, of course. Note that $H_{\!_{M,N}}$
commutes with $ \pbos$ so $ e^{-\alpha n H_{\!_{M,N}}} \pbos$ is
self-adjoint and positive.  The multiplication operator $\xi $ is also
positive and we can write $ \xi e^{-\alpha n H_{\!_{M,N}}} \pbos = [
\xi e^{-\alpha H_{\!_{M,N}}} \pbos] e^{-\alpha(n-1) H_{\!_{M,N}}}
\pbos$.  H\"older's inequality for traces of positive operators states
that $\Tr A B \leq \{\Tr A^n\}^{1/n} \{\Tr B^{ n/(n-1) }
\}^{(n-1)/n}$, and therefore
\begin{equation}
\frac {\Tr\, [\xi e^{-\alpha n H_{\!_{M,N}}} \pbos]}{\Tr\, 
[e^{-\alpha n H_{\!_{M,N}}} \pbos]} \leq \left( \frac {\Tr\, 
[ (\xi e^{-\alpha H_{\!_{M,N}}} \pbos)^n]}{\Tr\, [e^{-\alpha n H_{\!_{M,N}}} 
\pbos]}\right)^{1/n}\,.
\end{equation}

Consider the  bigger projection $\widehat\pbos$, which symmetrizes only among
particles $4,5,\dots,M$. It commutes with $H_{\!_{M,N}}$ and also with $\xi$, and hence
$ e^{-\alpha H_{\!_{M,N}}} \pbos \leq   e^{-\alpha H_{\!_{M,N}}} \widehat\pbos$. Since
$\xi \geq 0$, this yields the upper bound
\begin{equation}
\left( \frac {\Tr\, [ (\xi e^{-\alpha H_{\!_{M,N}}} \pbos)^n]}{\Tr\,
 [e^{-\alpha n H_{\!_{M,N}}} \pbos]}\right)^{1/n} \leq \left( \frac {\Tr\, 
[ \widehat\pbos (\xi e^{-\alpha H_{\!_{M,N}}})^n]}{\Tr\, [e^{-\alpha n H_{\!_{M,N}}} 
\pbos]}\right)^{1/n}\,.
\end{equation}

We now claim that
\begin{equation}\label{toshow}
\Tr\, [ \widehat\pbos (\xi e^{-\alpha H_{\!_{M,N}}})^n] \leq \Tr_3\, 
(\xi e^{-\alpha h})^n \, \Tr_{M-3}\, [ e^{-\alpha n H_{M-3,N}}
\widehat\pbos] \,,
\end{equation}
where $\Tr_3$ and $\Tr_{M-3}$ denote the trace over the first 3 and
last $M-3$ particles, respectively. Taking the limit $n\to\infty$ this
proves (\ref{lemeq2}).

To show (\ref{toshow}), we write $H_{\!_{M,N}}= H_{\!_{3,N}}\otimes
\id_{\!_{M-3}} + \id_{_{3}}\otimes H_{\!_{M-3,N}} + W$, with $W$ denoting the
interaction between the first 3 and the last $M-3$ particles. Note
that $W\geq 0$.  Using the Trotter product formula, we first replace
each factor $e^{-\alpha H_{\!_{M,N}}}$ by $(e^{-\alpha
H_{3,N}/m}e^{-\alpha (H_{M-3,N}+W)/m})^m$ for some integer $m$.  (Here
we abuse the notation slightly, omitting to write tensor products and
identity operators.)  For $\bfx=(\x_1,\x_2,\x_3)$, let $k(\bfx,\bfx')$
denote the integral kernel of $e^{-\alpha H_{3,N}/m}$.  Denoting by
$W_{\bfx}$ the multiplication operator on the subspace of the last
$M-3$ particles obtained by fixing the first $3$ to have positions
$\bfx$, and introducing $nm$ integration variables $\bfx_{ij}$, $1\leq
i\leq n$, $1\leq j \leq m$, we can write
\begin{multline}
  \Tr\, \left[ \widehat\pbos \left(\xi \left(e^{-\alpha
          H_{3,N}/m}e^{-\alpha (H_{M-3,N}+W)/m}\right)^m\right)^n
  \right] = \\ \int \prod_{ij} d\bfx_{ij} \prod_i \xi(\bfx_{i1})
  \prod_{ij} k(\bfx_{ij},\bfx_{i(j+1)}) \, \Tr_{M-3}\left[
    \widehat\pbos \prod_{i,j} e^{-\alpha (H_{M-3,N}+W_{\bfx_{ij}})/m}
  \right] \,,
\end{multline}
where we identify $\bfx_{i(m+1)}\equiv \bfx_{(i+1)1}$ and
$\bfx_{n(m+1)}\equiv\bfx_{1,1}$. By H\"older's inequality for traces,
we can estimate
\begin{align}\notag
  \biggl| \Tr_{M-3}\biggl[ \widehat\pbos \prod_{i,j} e^{-\alpha
        (H_{M-3,N}+W_{\bfx_{ij}})/m} \biggl] \biggl| &\leq \sup_{ij}
  \Tr_{M-3}\left[ \widehat\pbos e^{-\alpha n
      (H_{M-3,N}+W_{\bfx_{ij}})} \right] \\ &\leq \Tr_{M-3}\left[
    \widehat\pbos e^{-\alpha n H_{M-3,N}} \right]\,,
\end{align}
where in the last inequality we used the fact that $W_\bfx\geq 0$ and
that the partition function is monotone in the potential. By the
Feynman-Kac-It{\^o} formula~\cite[Sect.~15]{simon}, the integral
kernel $k(\bfx,\bfx')$ is bounded in absolute value by the kernel of
$e^{-\alpha h/m}$. Using this estimate and rewriting the integrals as
a trace we obtain
\begin{equation}\label{eq87}
\Tr\, \left[ \widehat\pbos \left(\xi \left(e^{-\alpha H_{3,N}/m}
e^{-\alpha (H_{M-3,N}+W)/m}\right)^m\right)^n \right] \leq \Tr_3\, 
(\xi e^{-\alpha h})^n \, \Tr_{M-3}\, [ e^{-\alpha n H_{M-3,N}}\widehat\pbos]\,.
\end{equation}
Letting $m\to\infty$ this yields (\ref{toshow}).
\end{proof}

We now use Lemma~\ref{3plemma} to obtain a bound on the various terms
in (\ref{mainineq}) and (\ref{22a}).  Lemma~\ref{3plemma} immediately
implies that
\begin{equation}
\big\langle |\x_1|^2 \big\rangle_\Bb \leq  e^{3\alpha Z} 
\| |x| e^{\alpha(\Delta-V(x))}|\x| \|_\infty \,,
\end{equation}
with $\|\,\cdot\,\|_\infty$ denoting operator norm. For positive
operators, the operator norm is bounded by the trace, in this case
given by $\Tr |x|^2 e^{\alpha(\Delta-V(x))}$. This expression, in
turn, is bounded for $\alpha$ large enough, as shown in the appendix.
In exactly the same way we can bound $\langle |\x_1|^4\rangle_\Bb $.

Moreover, we have that
\begin{equation}\label{boundwe}
\big\langle w_R(x_1-x_2) \big\rangle_\Bb \leq e^{3\alpha Z} \| \sqrt{w_R} 
e^{-\alpha h} \sqrt{w_R}\|_\infty \leq e^{3\alpha Z} 
\frac 1{(4\pi\alpha)^{3/2}} \int_{\R^3} w_R(x) dx \,.
\end{equation}
The last inequality can be seen as follows. Denote by $k(x,x')$ the
kernel of $e^{\alpha(\Delta-V(x))}$. The Feynman-Kac formula implies that 
$k(x,x')\leq (4\pi\alpha)^{-3/2}$ for any positive $V(x)$. Hence, for
any function $f\in L^2(\R^6)$,
\begin{align}\notag
 & \int dx dx' dy dy' \overline{f(x,y)} \sqrt{w_R(x-y)} k(x,x') k(y,y')
  \sqrt{w_R(x'-y')} f(x,y) \\ \notag &\leq (4\pi\alpha)^{-3/2} \int dx dx'
  k(x,x') \left[\int dy \sqrt{w_R(x-y)} |f(x,y)|\right] \left[\int dy
    \sqrt{w_R(x'-y)} |f(x',y)|\right] \\ &\leq (4\pi\alpha)^{-3/2}
  \left( \int w_R \right) \int dx dx' k(x,x') \left[\int dy |f(x,y)|^2
  \right]^{1/2} \left[\int dy |f(x',y)|^2 \right]^{1/2}\,,
\end{align}
where we used Schwarz's inequality in the last step.  The result now
follows from the fact that $e^{\alpha(\Delta-V(x))}\leq \id$.

Similarly, repeating the above argument with $\x_2$ in place of $\x$
and $(\x_1,\x_3)$ in place of $\y$, we obtain
\begin{align}\notag
  \big\langle U_R(\x_1-\x_2) \theta(2R-|\x_2-\x_3|) \big\rangle_\Bb &\leq
  e^{3\alpha Z} \frac 1{(4\pi\alpha)^3} \int_{\R^3} U_R(\x) d\x
  \int_{\R^3} \theta(2R-|\x|) d\x \\ &= e^{3\alpha Z} \frac 1{\alpha^3}
  \frac {2}{3\pi}R^3\,.
\end{align}
This finishes our bounds on the various terms appearing in (\ref{mainineq}) and
(\ref{22a}).

\subsection*{STEP 5. Collection  of All the  Terms and the Final Inequality.}

In this section we concatenate the various pieces of the lower bound
to the energy $E_0(M,N)$ in (\ref{mainineq}), and finish the proof of
Proposition~\ref{mainlemma}.  Inequality (\ref{mainineq}) contains
several terms.  All except $\infspec K$ were bounded in Step 4 and in
(\ref{22a}).  The essence of Step 3 is the bound on the main term
\begin{equation}\label{inspec}
\infspec K \geq \inf_\vecz \infspec
  U(\vecz) \geq  \inf_{\Phi} \E[\Phi] \, ,
\end{equation}
where $ \E[\Phi]$ is defined in (\ref{ephi2}).

Let us begin by disposing of the terms mentioned in Step 4. As shown
there, $\langle |x_1|^2\rangle_{\Bb}\leq \const$ and $\langle
|x_1|^4\rangle_{\Bb}\leq \const$ for some constant depending only on
$Z$. (Recall that $Z$ is a fixed number of order $1$.) Moreover, from
(\ref{boundwe}) and (\ref{wrint}) we see that (recalling that $M\leq N$) 
\begin{equation}
\frac {M^2 a }{N^2 \eps} \big\langle
  w_R(\x_1-\x_2)\big\rangle_\Bb \leq \const \frac a\eps \frac{R^2}{s^2} \,.
\end{equation}
This term will thus be negligible, if $R\to 0$ as $N\to\infty$
(keeping $\eps$ and $s$ fixed for the moment).  We are free to choose
the dependence of $R$ on $N$, and we choose $R$ to satisfy
\begin{equation}\label{R}
N^{-1/3}\gg R \gg N^{-2/3} \quad \mathrm{ as} \quad  N \to \infty \ . 
\end{equation}
The last term to estimate is then
\begin{equation}
\frac{a M^3}{N^2} \big\langle
  U_R(\x_1-\x_2) \theta(2R-|\x_2-\x_3|)\big\rangle_\Bb \leq \const a N R^3 \ll 1\,.
\end{equation}
Hence it follows from (\ref{mainineq}) and (\ref{22a}) that, for any
fixed $s$, $\eps$ and $\eta$ (recalling that $\lambda = \lim_{N\to
\infty} M/N$),
\begin{equation}\label{mainineq2}
\limi_{N\to\infty} \frac 1N (1+4\eta) E_0(M,N) \\ 
\geq \limi_{N\to\infty} \frac 1N \inf_\Phi \E[\Phi] + \lambda \kappa(\eta) 
- \const \lambda \eta \,.
\end{equation}

The only thing left is the minimization of $ \E[\Phi]$ given in
(\ref{ephi2}), which contains the numbers $D_1$, $D_2$ and $D_3$ in
(\ref{d1def})--(\ref{d3def}).  To evaluate them as $N\to \infty$ we
note that $\|W\|_1 \leq 4\pi a/N$, and $\|W\|_\infty \ll N$ for our
choice of $R$ in (\ref{R}).  Hence, we see that
\begin{equation}
\lim_{\delta\to 0} \lim_{J\to\infty} \lim_{N\to\infty} D_1 = 1 \ ,\ 
 \lim_{J\to\infty} \lim_{N\to\infty} D_2 = 0 \ , \ {\rm and} \
\lim_{N\to\infty} \frac 1 N D_3=0 \,.
\end{equation}
Using the fact that both $\|\nabla\Phi\|_2^2$ and $\|\Phi\|_2^2$ are
bounded relative to $\langle\Phi|K_0|\Phi\rangle$, and rescaling
$\Phi \to M^{1/2} \Phi$, we obtain
\begin{align}
  &\limi_{J\to\infty} \limi_{N\to\infty} \frac 1N \inf_\Phi \E[\Phi]  \geq \\
  &\limi_{R\to 0} \inf_\Phi \left\{ \lambda \langle
    \Phi|K_0|\Phi\rangle + (1-\eps) a \lambda^2 \int |\Phi(x)|^2
    |\Phi(y)|^2 U_R(\x-\y) dx dy + C \lambda
    \left(\|\Phi\|_2^2-1\right)^2\right\} \,. \notag
\end{align}

Note that the infimum can obviously be restricted to a set of bounded
$\langle \Phi|K_0|\Phi\rangle$, independent of $R$, since $U_R\geq 0$. Since $K_0\geq
-\eta \Delta$ this implies that we can assume that $\|\nabla\Phi\|_2$
is bounded independent of $R$, and hence also $\|\Phi\|_6$ by
Sobolev's inequality. Using the inequality (proved below)
\begin{equation}\label{usi}
\left| \int |\Phi(x)|^2 |\Phi(y)|^2 U_R(\x-\y) dx dy - 
4\pi \|\Phi\|_4^4 \right| \leq 8\pi R  \|\Phi\|_6^3 \|\nabla\Phi\|_2 \,,
\end{equation}
we see that we can interchange the limit and the infimum and thus obtain
\begin{equation}\label{ref2}
  \limi_{J\to\infty} \limi_{N\to\infty} \frac 1N  \inf_\Phi \E[\Phi]  \geq 
\inf_\Phi \left\{ \lambda \langle
    \Phi|K_0|\Phi\rangle + (1-\eps) 4\pi a \lambda^2 \|\Phi\|_4^2 + C
    \lambda \left(\|\Phi\|_2^2-1\right)^2\right\} \,.
\end{equation}
Ineq. (\ref{usi}) can be obtained in the following way. Using
Schwarz's inequality, as well as $\int U_R(y) dy = 4\pi$,
\begin{align}\notag
 &\left| \int |\Phi(x)|^2 |\Phi(y)|^2 U_R(\x-\y) dx dy - 4\pi
    \|\Phi\|_4^4 \right| \\ \notag &\leq \int dy U_R(y) \int dx |\Phi(x)|^2
  \left( |\Phi(\x)| + |\Phi(\x+\y)|\right) \left| |\Phi(\x)| -
    |\Phi(\x+\y)|\right| \\ &\leq 2\|\Phi\|_6^3 \int dy U_R(y) \left(
    \int dx \left |\Phi(x)| - |\Phi(x+y)|\right|^2 \right)^{1/2} \,.
\end{align}
The result now follows from the fact that $\| |\Phi| - |\Phi(\,\cdot\,
+y)|\|_2\leq |y| \|\nabla\Phi\|_2$, which can be seen by evaluating
the norm in Fourier space, using $|1-e^{-ip\cdot y}|^2 \leq |y|^2 |p|^2$, 
and also using the fact  that $\int U_R(y) |y| dy \leq R \int U_R(y) dy = 4\pi
R$.

Now, letting $C\to\infty$, we infer from (\ref{ref2}) that 
\begin{equation}\label{eqs}
\limi_{C\to\infty}\limi_{J\to\infty} \limi_{N\to\infty} \frac 1N  
\inf_\Phi \E[\Phi]  \geq \inf_{ \|\Phi\|_2=1} 
\left\{ \lambda \langle \Phi|K_0|\Phi\rangle + (1-\eps) 4\pi a 
\lambda^2 \|\Phi\|_4^2 \right\} \,.
\end{equation}
The final step is to remove the momentum cutoff in $K_0$, i.e., to let
$s\to 0$ in Eq. (\ref{eqs}).  Again, we claim that we can interchange
the limit and the infimum, at least to obtain a lower bound. Let
$\Phi_s$ denote a minimizer of the functional on the right side of
(\ref{eqs}).  Since $K_0\geq -\eta\Delta+V(x)$ and $V(x)\to \infty$ as
$|x|\to\infty$, a sequence $\Phi_{s_j}$ with $s_j\to 0$ as
$j\to\infty$ lies in a compact subset of $L^2(\R^3)$, and hence 
there exists a subsequence which converges strongly and pointwise
almost everywhere \cite{analysis} (both in $p$-space and $x$-space) to a function
$\Phi_0$ as $j\to \infty$, with $\|\Phi_0\|_2=1$.  All the $s$-independent
terms in the functional on the right side of (\ref{eqs}) are weakly
lower semicontinous. Moreover, by Fatou's Lemma \cite{analysis},
\begin{equation}\label{100}
\limi_{s\to 0} \int p^2 \big( 1-\chi_s(p)^2\big) 
|\widehat \Phi_s(p)|^2 dp \geq  \int p^2 |\widehat \Phi_0(p)|^2 dp\,.
\end{equation}
Hence the infimum and the limit $s\to 0$ can be interchanged for a
lower bound. In combination with  inequalities (\ref{eqs}) and (\ref{mainineq2}), we find that
\begin{multline}\label{eqs2}
\limi_{N\to\infty} \frac 1N (1+4\eta) E_0(M,N) \\ \geq  \inf_{ \|\Phi\|_2=1}
 \left\{ \lambda \left\langle \Phi\left| 
- \Delta + 2 p \cdot A(\x) +A(\x)^2 + V(\x) \right|\Phi\right\rangle + (1-\eps)
 4\pi a \lambda^2 \|\Phi\|_4^2 \right\} - \const \lambda\eta \,.
\end{multline}
(For a lower bound we simply dropped the positive terms $-2\eta\Delta$
and $\eta |x|^4$ in $K_0$.)  By letting $\eta\to 0$ and $\eps\to 0$
Proposition \ref{mainlemma} is proved. As explained in Step 1, this proves
Theorem \ref{energy}.

\medskip
{\it Remark\/} about the optimal choice of the parameters: In
Eq. (\ref{R}) we showed how the parameter $R$ has to depend on $N$, as
$N\to\infty$, in order to obtain the correct limit for the energy. The
explicit dependence on $N$ of the other parameters $J$, $C$, $s$,
$\eta$ and $\eps$ need not be specified so closely (unless we wish to
obtain a detailed error estimate). It suffices to let $J\to\infty$,
$C\to\infty$, $s\to 0$, $\eta\to 0$ and $\eps\to 0$ (in this order)
after taking the $N\to\infty$ limit. 


\section{Proof of Theorem \ref{condensation}} \label{sec3}

\subsection*{STEP 1. Proof of Part (i).} 
The fact that $\Gamma$ is a convex set follows easily from its
definition.  Namely, if $\gamma_{\!_N}$ and $\bar\gamma_{\!_N}$ are
two approximate ground state sequences, and $0\leq \lambda\leq 1$,
then $\lambda \gamma_{\!_N} + (1-\lambda)\bar\gamma_{\!_N}$ is
certainly also an approximate ground state sequence, whose reduced one
particle density matrix is given by
$\lambda\gamma_{\!_N}^{(1)}+(1-\lambda)\bar\gamma_{\!_N}^{(1)}$.

Compactness of $\Gamma$ is also not difficult to see. Given a sequence
$\gamma_i\in \Gamma$, the Banach-Alaoglu Theorem implies the existence
of a subsequence such that $\gamma_i\rightharpoonup \gamma_\infty$ for
some $\gamma_\infty$ in the weak-* sense as $i\to\infty$.  As already
remarked in the introduction, the fact that $\Tr\, H_0 \gamma_i \leq
\const$ implies that $\gamma_i \to \gamma_\infty$ in trace norm. To
prove compactness we have to show that $\gamma_\infty\in \Gamma$.

By definition, corresponding to every $\gamma_i$ there is an
approximate ground state sequence $\gamma_{\!_{N,i}}$. That is, there
is a number $N_i$ such that $N\geq N_i$ implies that
$\|\gamma_{\!_{N,i}}^{(1)}-\gamma_i\|\leq 1/i$ and $| N^{-1}\Tr\,
H_{\!_N} \gamma_{\!_{N,i}} - E^{\rm GP}(a)|\leq 1/i$. (Here,
$\|\,\cdot\,\|$ denotes trace norm.) We can assume that $N_i\to\infty$
as $i\to\infty$. Now, for given $N$, let $\hat \imath(N)$ be the largest integer $i$
such that $N\geq N_i$. Then $\hat \imath(N)\to \infty$ as $N\to
\infty$, and hence the sequence $\gamma_{\!_{N,\hat \imath(N)}}$ is an
approximate ground state sequence.  Moreover, $\|\gamma_{\!_{N,\hat
\imath(N)}}^{(1)}-\gamma_\infty\|\leq
\|\gamma_{\!_{N,\hat \imath(N)}}^{(1)}-\gamma_{_{\hat \imath(N)}}\|+ \|\gamma_{_{\hat
  \imath(N)}}-\gamma_\infty\|\to 0$ as $N\to\infty$. This proves that
$\gamma_\infty\in \Gamma$, and hence $\Gamma$ is compact.

\subsection*{STEP 2. An Extension of Theorem~\ref{energy}.}

A key step in the proof of Theorem~\ref{condensation} is an extension
of the lower bound in Theorem~\ref{energy} to the case of a perturbed
Hamiltonian, where we replace the one-particle part $H_0$ of the
Hamiltonian (\ref{ham}) by $H_0+\s$, where $\s$ is a bounded hermitian
operator on the one-particle space $L^2(\R^3)$. Let $H_{\!_N}^{(\s)}$
denote the perturbed $N$-particle operator 
$$H_{\!_N}^{(\s)}=H_{\!_N} +
\sum_{i=1}^N \s^{(i)}\, ,
$$
and let $E_0^{(\s)}(N)=\infspec H_{\!_N}^{(\s)}$ denote its ground
state energy.  Correspondingly, define the perturbed GP functional
$\Egp_{(\s)}$ as in (\ref{gpenergy}), with $H_0+\s$ in place of $H_0$,
and let $E_{(\s)}^{\rm GP}(a)$ denote its infimum over all $\phi$ with
$\|\phi\|_2=1$. Then we have the following extension of
Theorem~\ref{energy}, to whose proof we will devote the remainder of
this subsection.

\begin{prop}\label{thm1ext}
For all bounded hermitian operators $\s$,
\begin{equation}\label{eee}
\liminf_{N\to\infty} \frac 1N E_0^{(\s)}(N) \geq E_{(\s)}^{\rm GP}(a)\,.
\end{equation}
\end{prop}

We start by noting that in order to prove Proposition~\ref{thm1ext} it
suffices to prove it in the special case in which $\s$ is a finite
rank operator with exponentially decaying eigenfunctions. In
particular, we can assume that its integral kernel $\s(x,y)$ satisfies
a bound
\begin{equation}\label{akern}
|\s(x,y)|\leq B \exp\left(- D(|x|+|y|)\right)
\end{equation}
for some positive constants $B$ and $D$. This can be seen as follows.
Let $\{f_i\}_{i=1}^\infty$ be an orthonormal basis for $L^2(\R^3)$ such
that $|f_i(x)| < B_i\exp(-D_i|x|)$ for some choice of constants
$B_i,\, D_i>0$ and let $P_n$ denote the projection onto the first $n$
of these functions. Clearly, $P_n \to \id$ strongly as $n\to \infty$.  Then,
for any bounded $\s$, $P_n \s P_n$ is of the desired form, i.e., it
has finite rank and its integral kernel satisfies a bound of the form
(\ref{akern}). For any one-particle density matrix $\gamma$,
\begin{equation}\label{normm}
\Big| \Tr\big[\gamma(\s-P_n \s P_n)\big] \Big| \leq 
\, \left\| \frac 1{\sqrt{H_0}}\big(\s-P_n\s P_n\big) \frac 1{\sqrt{H_0}}\right\|
\ \Tr [H_0\gamma] \,,
\end{equation}
with $\|\,\cdot\,\|$ denoting operator norm.  
Since $H_0^{-1/2}$ is compact and, therefore, is the norm limit of
finite rank operators, it is easy to see that the norm in
(\ref{normm}) goes to zero as $n\to\infty$.  On the other hand the set
of numbers $\Tr [H_0\gamma]$ that arise from those $\gamma$'s that
come from approximate ground states is bounded. Consequently, both
sides of (\ref{eee}) can be approximated to within any desired
$\varepsilon$ by replacing $\s$ by $P_n\s P_n$ and choosing $n$ large
enough --- which implies the statement.

Thus we can assume (\ref{akern}) henceforth. The proof of
Proposition~\ref{thm1ext} then follows exactly the same lines as the
proof of Theorem~\ref{energy}. In fact, our proof of
Theorem~\ref{energy} has the advantage of being almost completely
independent of the exact form of the Hamiltonian. The only place where
we used the explicit form is Lemma~\ref{3plemma}, which was used to
bound expectation values of certain one-, two- and three-body
operators in the zero-temperature state of $H_{\!_{M,N}}$. We now have
to bound the expectation value of these operators in the
zero-temperature state of $H_{\!_{M,N}}^{(\s)}$, which we
denote as $\langle\,\cdot\,\rangle_\Bb^{(\s)}$. (Here, the operator
$H_{\!_{M,N}}^{(\s)}$ is defined in the obvious way. Its ground
state energy will be denoted by $E_0^{(\s)}(M,N)$.)  To this end,
Lemma~\ref{3plemma} can be extended in the following way.

\begin{lem}\label{3plemmam}
  Let $\xi (\x_1,\x_2,\x_3)$ be any positive function of $\x_1$,
  $\x_2$ and $\x_3\in\R^3$. Let $\widehat\s$ denote the rank one
  operator on the one-particle space with integral kernel given by the
  right side of (\ref{akern}).  With $V=$ the one-body potential
  appearing in $H_{\!_{M,N}}$, we define the three-body, independent
  particle Hamiltonian
\begin{equation}\label{h3mod}
h^{(\s)}= -\Delta_1-\Delta_2-\Delta_3 +
  V(\x_1) +V(\x_2)+V(\x_3) - \widehat \s_1 -\widehat \s_2 -\widehat \s_3 \,.
\end{equation}
  Let $\alpha>0$ and let
  $e^{-\alpha h^{(\s)}}(\x_1,\x_2,\x_3\ ; \ \y_1,\y_2,\y_3)$ be the `heat
  kernel' of $h^{(\s)}$ at `inverse temperature' $\alpha$. Finally, consider
  the modified integral kernel 
\begin{equation}
e^{-\alpha h^{(\s)}}(\x_1,\x_2,\x_3\ ; \ \y_1,\y_2,\y_3)
  \sqrt{\xi(\x_1,\x_2,\x_3) \xi( \y_1,\y_2,\y_3)}
\end{equation} 
and let $\Lambda^{(\s)}$ denote its largest eigenvalue (i.e., its norm
as a map from $L^2(\R^9)$ to $L^2(\R^9)$).  Then
\begin{equation}\label{lemeq}
\big\langle \xi(\x_1,\x_2,\x_3) \big\rangle_\Bb^{(\s)} \leq \Lambda^{(\s)} 
\exp\{\alpha(E_0^{(\s)}(M,N)-E_0^{(\s)}(M-3,N))\}\, .
\end{equation}
\end{lem}

The proof follows along the same lines as the proof of
Lemma~\ref{3plemma}, except for one step. Before Eq. (\ref{eq87}), it
was necessary to get an upper bound on the absolute value of the
integral kernel of $\exp\{-\alpha H_{\!_{3,N}} /m\}$ in terms of the kernel
of $\exp\{-\alpha h/m\}$, which can be obtained with the help of the
Feynman-Kac-It{\^o} formula. In the case considered here, we need an
upper bound on the integral kernel of $\exp\{-\alpha H_{\!_{3,N}}^{(\s)}/m\}$.
We will now show that the absolute value of this kernel is bounded
above by the kernel of $\exp \{-\alpha h^{(\s)}/m\}$ for the modified
three-particle operator $h^{(\s)}$ in (\ref{h3mod}).

This claim follows from the Trotter product formula, together with the
Feynman-Kac-It{\^o} formula, in the following way. Since $\s$ is a
bounded (in fact, finite rank) operator, we can write
\begin{equation}
e^{-\alpha H_{3,N}^{(\s)}/m} = \lim_{n\to\infty} \left[ e^{-\alpha H_{3,N} /n} 
\left(1 - \frac {\alpha}{n} \s\right) \right]^{n/m}\,.
\end{equation}
By the Feynman-Kac-It{\^o} formula, (\ref{akern}) and the definition of $\widehat \s$, 
\begin{equation}
\left| \left[ e^{-\alpha H_{3,N} /n} \left(1 - \frac {\alpha}{n} \s\right) 
\right]^{n/m}(x,y)\right| 
\leq  \left[ e^{-\alpha h /n} \left(1 + \frac {\alpha}{n} \widehat \s\right) 
\right]^{n/m} (x,y)\,.
\end{equation}
In the limit $n\to\infty$, the operator on the right side converges
strongly to $e^{-\alpha h^{(\s)} /m}$. This proves our claim.

For the application of this Lemma, as in Section 2, Step 4, it is
necessary to have some bounds on the kernel of $e^{-\alpha h^{(\s)}}$.
In particular, we need that the kernel is bounded, and that its
diagonal decays for large $|x|$ at least like $|x|^{-\const \alpha}$
for some positive constant. As for the case $\s=0$, these properties
are again shown in the appendix. It is there that the exponential
decay of the kernel of $\widehat\s$ gets used.

As already mentioned, except for the replacement of
Lemma~\ref{3plemma} by Lemma~\ref{3plemmam}, the proof of
Proposition~\ref{thm1ext} consists of simply mimicking the discussion
of the proof of Theorem~\ref{energy} given in Section~\ref{sec2}.

\subsection*{STEP 3. $\Gamma$ Contains Projections onto GP Minimizers.}

Let $\Phi^{\rm GP}\subset L^2(\R^3)$ denote the {\it set\/} of all minimizers
of the GP functional (\ref{gpenergy}). We now consider the special case where
$\s=-|\phi\rangle\langle\phi|$ for some $\phi\in \Phi^{\rm GP}$. In this case, we claim that
\begin{equation}\label{thg}
\lim_{N\to\infty} \frac 1N E_0^{(\lambda \s)}(N) = E^{\rm GP}(a) -\lambda 
\end{equation}
for any $\lambda\geq 0$. Given Theorem~\ref{energy}, the lower bound
is trivial in this case. The upper bound can be derived in the same
way as the upper bound for Theorem~\ref{energy} in \cite{S03}. The
arguments there also apply to this case, and the expectation value of
$\s$ in the trial state can easily be estimated using the methods in
\cite{dyson,LSY00}. (In the non-rotating case, this was carried out in~\cite{RSthesis}.)

Taking the derivative of (\ref{thg}) at $\lambda>0$, Griffiths'
argument \cite {Griffiths, LSYjust} implies that the one-particle
density matrix of a ground state of $H_{\!_N}^{(\lambda \s)}$ converges
to $|\phi\rangle\langle\phi|$ as $N\to\infty$ in this case. Hence, by
a similar \lq diagonal\rq\ argument as at the end of the proof of part
(i) of Theorem~\ref{condensation}, we can find a sequence
$\lambda_{\!_N}$ with $\lambda_{\!_N}\to 0$ as $N\to\infty$ such that
the ground state of $H_{\!_N}^{(\lambda_{\!_N}\! \s)}$ represents an
approximate ground state sequence for the $\lambda=0$ problem, and its
reduced one-particle density matrix converges to
$|\phi\rangle\langle\phi|$ as $N\to\infty$. This shows that
$|\phi\rangle\langle\phi|\in \Gamma$ for any $\phi\in\Phi^{\rm GP}$.

(Remark: The claim of this subsection can in principle be proved by
simply constructing an appropriate approximate ground state. However,
although the one-particle density matrix of the trial state used in
\cite{S03} converges to $|\phi\rangle\langle\phi|$ as $N\to\infty$,
this does not immediately imply that $|\phi\rangle\langle\phi|\in
\Gamma$ since the trial state is not symmetric! This explains the
somewhat different reasoning in this subsection.)

We note that also $|\phi\rangle\langle\phi|\in \Gamma_{\rm ext}$ for
all $\phi\in\Phi^{\rm GP}$. This follows from the fact that all
elements of $\Gamma$ are positive operators, and a rank one operator
cannot be written as a non-trivial convex combination of two positive
operators. In the next subsection, we will show that all elements of
$\Gamma_{\rm ext}$ are of the form $|\phi\rangle\langle\phi|$ with
$\phi\in \Phi^{\rm GP}$. 

\subsection*{STEP 4. Proof of Parts (ii) and (iii).}

For a given $\gamma\in \Gamma$, let $\gamma_{\!_N}$ be an approximate
ground state sequence for $H_{\!_N}$, with $\gamma_{\!_N}^{(1)}\to
\gamma$ as $N\to\infty$. By Proposition~\ref{thm1ext} we have that,
for any bounded hermitian operator $\s$ and any $\lambda\in \R$,
\begin{equation}
E^{\rm GP}(a) + \lambda\, \Tr\, \s\gamma = \lim_{N\to\infty} \frac 1N 
\Tr\, H_{\!_N}^{(\lambda \s)}\gamma_{\!_N} \geq E^{\rm GP}_{(\lambda \s)}(a)\,.
\end{equation}
Upon dividing by $\lambda$ and letting $\lambda\to 0$, this yields
\begin{equation}
 \Tr\, \s\gamma \geq \lim_{\lambda\searrow 0} 
\frac {E^{\rm GP}_{(\lambda \s)}(a) - E^{\rm GP}(a)}{\lambda}\,.
\end{equation}
 We claim that
\begin{equation}\label{minc}
 \lim_{\lambda\searrow 0} 
\frac {E^{\rm GP}_{(\lambda \s)}(a) - E^{\rm GP}(a)}{\lambda} 
= \min_{\phi\in\Phi^{\rm GP}} \langle \phi|\s|\phi\rangle\,.
\end{equation}
Using $\phi\in \Phi^{\rm GP}$ as a trial function, we immediately see
that $E^{\rm GP}_{(\lambda \s)}(a)\leq E^{\rm GP}(a)+ \lambda
\langle\phi|\s|\phi\rangle$ for all $\phi\in \Phi^{\rm GP}$. For the
other direction, we use a minimizer of $\Egp_{(\lambda \s)}$ as a
trial state for $\Egp$. As $\lambda\to 0$, this sequence of minimizers
will have a subsequence that converges strongly to a minimizer of
$\Egp$.  Hence, for some $\phi\in\Phi^{\rm GP}$, $
\lim_{\lambda\searrow 0} \lambda^{-1} (E^{\rm GP}_{(\lambda \s)}(a) -
E^{\rm GP}(a)) \geq \langle \phi|\s|\phi\rangle$, which proves our
claim. (Note that this argument also proves that the right side of
(\ref{minc}) is a true minimum and not merely an infimum.)

We have thus shown that, for every bounded hermitian operator $\s$, and
every $\gamma\in \Gamma$,
\begin{equation}\label{key}
 \Tr\, \s\gamma \geq \min_{\phi\in\Phi^{\rm GP}} \langle \phi|\s|\phi\rangle\,.
\end{equation}
Replacing $\s$ by $-\s$, this also implies that $\Tr\, \s\gamma \leq
\max_{\phi\in\Phi^{\rm GP}} \langle \phi|\s|\phi\rangle$. Inequality
(\ref{key}) is the key to the proof of statements (ii) and (iii) in
Theorem~\ref{condensation}.

Let $P_n$ be a rank $n$ projection, and let $P_n\Gamma=\{P_n\gamma
P_n\, : \, \gamma\in\Gamma\}$. When $\gamma$ is a bounded operator
on $L^2(\R^3)$, $P_n\gamma P_n$ can be identified with an $n\times n$
complex matrix, and hence with a vector in $\R^{2n^2}$. We make this
identification (denoted by $\iota$ in the following) in order to be
able to use finite-dimensional convexity theory (see, e.g.,
\cite{rockafellar}).  Note that $\iota$ is linear and continuous, and
hence the set $B_n=\iota P_n\Gamma=\{\iota P_n\gamma P_n\, : \,
\gamma\in\Gamma\}$ is a closed convex subset of $\R^{2n^2}$. An {\it
  exposed point\/} \cite{rockafellar} of a convex set $\mathcal{ C }
\subset \R^m$ is an extreme point $p$ of $\mathcal{C}$ with the additional
property that there is a tangent plane to $\mathcal{C}$ containing $p$ but
containing no other point of $\mathcal{C}$.  (For an example of points
that are extreme but not exposed, let $\mathcal{ C} \subset \R^2$ be a
square with each corner rounded off into a quarter of a circle.  The
extreme points are all the points on the four quarter-circles,
including their endpoints, but the endpoints are not exposed.)

An equivalent way to say this is that an exposed point $p$ in
$\mathcal{C} \subset \R^{m}$ is characterized by the existence of a
vector $a\in \R^{m}$ (a normal to the tangent plane) such that
\begin{equation} \label{normal}
(a,p)\leq (a,b) \quad {\rm for\ all\ } b\in \mathcal{C}\,,
\end{equation}
with equality {\it if and only if\/} $b=p$. (Here, $(\cdot\, , \, \cdot)$
denotes the standard inner product in $\R^{m}$.) 

For a fixed $n$, an exposed point of $B_n\subset \R^{2n^2}$
corresponds to some $P_n\widetilde\gamma P_n \in P_n \Gamma$. This density
matrix $\widetilde\gamma $ may not be unique and it may depend on $n$,
but this is of no concern to us. 

We note that our space of density matrices is a complex space and,
therefore, we have to translate (\ref{normal}) to this setting. For
any two bounded operators $\gamma, \, \gamma'$ (not necessarily in
$\Gamma$), the real inner product $(\cdot\, , \, \cdot)$ becomes
\begin{equation}
(\iota P_n \gamma P_n,\, \iota  P_n \gamma' P_n) = 
\Re \, \Tr (P_n \gamma^\dagger P_n \gamma') \, ,
\end{equation}
where $ \gamma^\dagger$ is the adjoint of $\gamma$. 
Translated to our
original space, this means that if $P_n\widetilde\gamma P_n$ is an exposed
point of $P_n\Gamma$, then there exists an operator $\s$ (with $P_n\s
P_n=\s$) such that
\begin{equation}
\Re\, \Tr\, \s \widetilde\gamma \leq {\rm \Re\,} \Tr\, \s \gamma   
\quad {\rm for\ all\ }\gamma\in\Gamma\,,
\end{equation}
or, equivalently, there exists a {\it hermitian\/} $\s$ such that 
\begin{equation}\label{tran}
 \Tr\, \s \widetilde\gamma \leq  \Tr\, \s \gamma   \quad {\rm for\ all\ }\gamma\in\Gamma\,.
\end{equation}
Note that, by definition, equality holds in (\ref{tran}) if and only
if $P_n\widetilde\gamma P_n = P_n\gamma P_n$.  We now use inequality
(\ref{tran}), with $\gamma=|\phi\rangle\langle\phi|$, where
$\phi\in\Phi^{\rm GP}$ minimizes $\langle\phi |\s|\phi\rangle$ among
all GP minimizers. We know from Step 3 that this $\gamma$ is an
element of $\Gamma$.  The inequalities (\ref{key}) (applied to
$\widetilde\gamma$) and (\ref{tran}) for this special choice of
$\gamma$ together imply that there is actually {\it equality\/} in
this case, and thus that $P_n\widetilde\gamma
P_n=P_n|\phi\rangle\langle\phi|P_n$. That is, all exposed points of
$P_n\Gamma$ are of the form $P_n |\phi\rangle\langle\phi| P_n$, with
$\phi\in\Phi^{\rm GP}$.

We can go further and conclude that {\it all} extreme points in
$P_n\Gamma$ are of this form, not only the exposed points.  This
follows from the fact that the set of GP minimizers is closed,
together with Straszewicz's Theorem \cite[Thm.~18.6]{rockafellar}
which states that the exposed points are a dense subset of the extreme
points.

Carath{\'e}odory's Theorem \cite[Thm.~17.1]{rockafellar} implies that
every $P_n\gamma P_n \in P_n \Gamma$ can be written as a convex combination of
$2n^2+1$ extreme points. That is, there exist $\lambda_i\geq 0$ with
$\sum_i{\lambda_i}=1$ such that
\begin{equation}\label{meas}
P_n\gamma P_n = P_n \left(  \sum_{i=1}^{2n^2+1} 
\lambda_i |\phi_i\rangle\langle\phi_i| \right) P_n \,,
\end{equation}
with $\phi_i\in\Phi^{\rm GP}$ for all $i$. This equation defines an
atomic (i.e., point) measure $d\mu_n(\phi)$ supported on the (compact)
space of projections onto GP minimizers. Let us provisionally call
this space $\Delta$, with the intention of showing that
$\Gamma_{\mathrm{ext}}=\Delta$.

For every $\psi$ with $P_n \psi =\psi$ we have thus shown that
\begin{equation} \label{meas2}
\langle \psi | \gamma | \psi \rangle = 
\int_\Delta d\mu_n(\phi) |\langle \psi | \phi\rangle|^2
\quad\quad \mathrm{with} \quad\quad \int_\Delta d\mu_n (\phi) =1\, .
\end{equation}
To complete the proof of Theorem~\ref{condensation} we wish to take
the limit $n\to \infty$ in (\ref{meas2}).  We choose $P_n$ in such a
way that $P_n$ converges strongly to the identity as $n\to \infty$.
The sequence $d\mu_n$ has a subsequence that converges weakly to some
measure $d\mu$ with $\int_\Delta d\mu =1$ (see \cite[vol. 1, Thm. 12.7 and
12.10]{choq}). This implies that, for $\psi$ in a dense subset of
$L^2(\R^3)$ (namely, those $\psi$ for which $P_n\psi =\psi$ for some
$n$), 
\begin{equation} \label{meas3}
\langle \psi | \gamma | \psi \rangle = 
\int_\Delta d\mu(\phi) |\langle \psi | \phi\rangle|^2
\quad\quad \mathrm{with} \quad\quad \int_\Delta d\mu (\phi) =1\, .
\end{equation}
Since (\ref{meas3})
holds for a dense set of $\psi$, it actually holds for all $\psi$ by continuity.
That is, $\gamma= \int_\Delta d\mu(\phi) | \phi\rangle \langle \phi|$
in the weak sense.

Note that there is a representation (\ref{meas3}) for $\gamma \in
\Gamma_{\rm ext}$ (since there is such a representation for all
$\gamma \in \Gamma$).  It is not hard to see that for an extreme
$\gamma$ the corresponding Borel measure $d\mu$ must be an atomic
measure at a single point in $\Delta$. Another way to say this is that
$\Gamma_{\rm ext} \subset \Delta$, which is exactly part (ii) of
Theorem~\ref{condensation} (since we have already proved in Step~3
that $\Delta \subset \Gamma_{\rm ext}$).

Part (iii) of Theorem~\ref{condensation} follows from (\ref{meas3}),
together with part (ii). This completes the proof of Theorem~\ref{condensation}.

\medskip 
We conclude with the direct proof of (\ref{finite}), which
was promised just after the statement of Theorem 2. We start with
(\ref{meas}) and choose $P_n$ to be the projection onto the largest
$n$ eigenvalues of $\gamma$, with $n$ large enough so that $\Tr\,
|\gamma -P_n \gamma P_n |< \varepsilon^2/8$.  We now denote $P_n=P$,
$1-P_n =Q$ and $B= \sum_i \lambda_i |\phi_i\rangle \langle \phi_i | $.
{F}rom (\ref{meas}) (and a little algebra) we learn that $\gamma -B
=Q(\gamma-B)Q -QBP -PBQ$. Thus $\Tr\, | \gamma -B| \leq \Tr\, (|Q\gamma Q| +
|QBQ| + 2|QBP|)$.  Obviously, $\Tr\, Q\gamma Q <\varepsilon^2/8$ and
since $\Tr\, B=\Tr\,\gamma=1$, we also have $\Tr\, |QBQ|=\Tr\, QBQ = \Tr\, (1-P)B
=\Tr\,(1-P)\gamma =\Tr\, Q\gamma Q <\varepsilon^2/8$. The remaining term
can be bounded, using Schwarz's inequality, by $(\Tr\, QBQ)^{1/2} (\Tr\,
PBP)^{1/2} < \varepsilon/\sqrt8$. This proves (\ref{finite}).



\appendix
\section*{Appendix: Heat Kernel Estimates}\label{heatapp}

In this appendix we derive an upper bound on the heat kernel for a
general Schr\"odinger operator.  This bound will show,  in particular,
that for any $s>0$ and $\alpha$ large enough (depending on $s$)
\begin{equation}\label{appf}
\Tr\, |x|^s e^{\alpha(\Delta-V)} < \infty
\end{equation}
if $V(x)\geq C_1 \ln(|x|) - C_2$ for some constants $C_1>0$ and $C_2$.
This property was used in the proof of Theorem~\ref{energy}.
(Actually, in the proof of Theorem~\ref{energy} we used only the cases
$s=2$ and $s=4$ (see Step 4 of Sect. \ref{sec2}) because we assumed
$A= \half \Omega \wedge x$, but (\ref{appf}) permits the inclusion of a magnetic
field with polynomial growth of $A$.)

Our bound on the heat kernel follows an idea of Symanzik \cite{sym}.
Using the Feynman-Kac formula for the integral kernel, we can write\
\begin{equation}
e^{\alpha(\Delta-V)}(x,y)  = 
\int d\mu_{x,y}(\omega) \exp \left(- \int_0^\alpha ds\, V(\omega(s))\right) \,,
\end{equation}
where $d\mu_{x,y}$ denotes the conditional Wiener measure for paths $\omega$ going
from $x$ to $y$ in time $\alpha$. By Jensen's inequality we have, for
any given path $\omega$,
\begin{equation}
\exp \left( -\int_0^\alpha ds\, V(\omega(s))\right) \leq \frac 1\alpha 
\int_0^\alpha ds\, \exp\left(-\alpha V(\omega(s))\right)\,.
\end{equation}
Therefore (using Fubini's Theorem) 
\begin{align}\label{fubb}
   e^{\alpha(\Delta-V)}(x,y) 
  & \leq \frac 1\alpha \int_0^\alpha ds\, \int d\mu_{x,y}(\omega)
  \exp\left(-\alpha V(\omega(s))\right) \notag \\ 
&= \frac 1\alpha
  \int_0^\alpha ds\, \left\{e^{s\Delta} e^{-\alpha V}
      e^{(\alpha-s)\Delta}\right\}(x,y)  \,.
\end{align}

To evaluate the trace in (\ref{appf}), we only need the heat kernel on
the diagonal, i.e., for $x=y$. The integral kernel of
$e^{t\Delta}$ is given by
\begin{equation}
j_t(x-y) \equiv \frac 1{(4\pi t)^{3/2}} e^{-|x-y|^2/4t}\, ,
\end{equation}
which leads to
\begin{equation}
\left\{e^{s\Delta} e^{-\alpha V}
      e^{(\alpha-s)\Delta}\right\}(x,x)
= \frac 1{(4\pi)^3} \frac 1{(t \alpha)^{3/2}} \int_{\R^3} dy\, 
e^{-\alpha V(y)} \exp\left( -|x-y|^2 / 4t\right) \, ,
\end{equation}
where  $t$ is defined by $1/t \equiv 1/s+1/(\alpha-s)$. 

Let us change the integration variable from $s$ to $t$, and introduce
the function
\begin{equation}
h_\alpha(x) = \frac 2\alpha \int_0^{\alpha/4} dt\, \frac 1{\sqrt{1-4t/\alpha}}\, 
j_t(x) \, .
\end{equation}
Then the bound (\ref{fubb}) yields
\begin{equation}
e^{\alpha(\Delta-V)} (x,x) \leq 
\frac 1{(4\pi \alpha)^{3/2}}  \left(e^{-\alpha V} * h_\alpha\right)(x) \,,
\end{equation}
with $*$ denoting convolution. Note that $\int h_\alpha(x) dx = 1$. It
is easy to see that $h_\alpha(x)\sim \exp(-|x|^2/\alpha)$ for large
$|x|$. Hence, if $V(x)$ increases logarithmically with
$|x|$, we see that the diagonal of the heat kernel decays at least as
$|x|^{-\const \alpha}$ for large $|x|$. Thus, we can choose $\alpha$
large enough to ensure that (\ref{appf}) is finite.

\medskip
For the proof of Theorem~\ref{condensation} it is necessary to extend this result to
the case where $-\Delta+V$ is replaced by $-\Delta+V+K$, with $K$ a
finite rank operator. As explained there, we can restrict ourselves to
the case when $K$ has exponentially decaying eigenfunctions. I.e., we
can assume that the kernel of $K$, which we denote by $K(x,y)$,
satisfies a bound
\begin{equation}\label{kkern}
K(x,y)\leq B e^{-D(|x|+|y|)}
\end{equation}
for some constants $B>0$ and $D>0$. Again we want to show that, for
any $s>0$ and $\alpha$ large enough (depending on $s$),
\begin{equation}\label{appf2}
\Tr\, |x|^s e^{\alpha(\Delta-V-K)} < \infty
\end{equation}
if $V(x)\geq C_1 \ln(|x|) - C_2$ for some constants $C_1>0$ and $C_2$.

With the notation $L_t=e^{t(\Delta-V)}$, we can use the
Dyson expansion to write
\begin{equation}\label{dysexp}
e^{\alpha(\Delta-V-K)} = L_\alpha + \sum_{n\geq 1} (-1)^n \int_{\sum_i t_i = \alpha} 
dt_0 dt_1 \cdots dt_n\, L_{t_0} K L_{t_1} K \cdots K L_{t_n} \,.
\end{equation}
We have already derived an upper bound on the kernel of $L_\alpha$
above. The kernel of the terms for $n\geq 1$ in the sum can be bounded
as follows. First of all, the Feynman-Kac formula tells us that 
since $V\geq 0$ we have the inequality
$L_t(x,y)\leq j_t(x-y)$ for the kernel of $L_t$. Moreover, using
(\ref{kkern}) and denoting by $\Phi$ the function
$\Phi(x)=\sqrt{B}e^{-D|x|}$, we have
\begin{equation}
\big|\left(L_{t_0} K L_{t_1} K \cdots K L_{t_n} \right)(x,y)\big|
\leq j_{t_0}* \Phi(x)\,  \prod_{i=1}^{n-1} \langle \Phi| L_{t_i} | \Phi\rangle 
\,j_{t_n}* \Phi(y) \,.
\end{equation}
Since $L_t\leq \id$, we have $\langle \Phi| L_{t_i} | \Phi\rangle \leq
\|\Phi\|_2^2$. Denoting
\begin{equation}
\xi_\alpha(x) = \|\Phi\|_2^{-1} \sup_{0<t<\alpha} j_t* \Phi(x) \,,
\end{equation}
we thus have
\begin{equation}
\big|\left( L_{t_0} K L_{t_1} K \cdots K L_{t_n} \right) (x,y)\big|
\leq \xi_\alpha(x) \xi_\alpha(y) \|\Phi\|_2^{2n}\,.
\end{equation}
The integral over the simplex in (\ref{dysexp}) yields a factor
$\alpha^n /n!$, and hence
\begin{equation}
\left| e^{\alpha(\Delta-V-K)} (x,y)\right| \leq  
e^{\alpha(\Delta-V)} (x,y) 
+ \left( e^{\alpha \|\Phi\|_2^2}-1\right)  \xi_\alpha(x) \xi_\alpha(y)\,.
\end{equation}
Since $\xi_\alpha$ decays exponentially for large $|x|$ this proves
our claim (\ref{appf2}).


\end{document}